\documentclass[twocolumn]{aastex631}

\usepackage[T1]{fontenc}
\usepackage{lmodern}

\usepackage{xspace}
\usepackage{amsmath}
\usepackage{amssymb}
\usepackage{url}
\usepackage{graphicx}
\usepackage{hyperref}
\usepackage{multirow}
\usepackage{tikz}

\newcommand{\phnm}{\phm{$-$}}
\newcommand{\phtnm}{\phm{\Large{$^a$}}}

\newcommand{\psrchive}{\textsc{psrchive}\xspace}
\newcommand{\pint}{\textsc{pint}\xspace}

\newcommand{\dmu}{\textrm{pc\,cm}$^{-3}$\xspace}
\newcommand{\rmu}{\textrm{rad\,m}$^{-2}$\xspace}

\newcommand{\Msun}{\textrm{M}$_\sun$\xspace}

\newcommand{\us}{$\mu$\textrm{s}\xspace}

\newcommand{\GHz}{\textrm{GHz}}
\newcommand{\ms}{\textrm{ms}}
\newcommand{\s}{\textrm{s}}

\newcommand{\RM}{\textrm{RM}}

\shortauthors{Fiore et al.}
\shorttitle{J1022+1001 Profile Variability}

\begin{document}

\title{Pulse Profile Variability of PSR J1022+1001 in NANOGrav Data}

\correspondingauthor{W.\ Fiore}
\email{william.fiore@nanograv.org}

\author[0000-0001-5645-5336]{William Fiore}
\affiliation{Department of Physics and Astronomy, University of British Columbia, 6224 Agricultural Road, Vancouver, BC V6T 1Z1, Canada}
\affiliation{Department of Physics and Astronomy, West Virginia University, P.O.\ Box 6315, Morgantown, WV 26506, USA}
\affiliation{Center for Gravitational Waves and Cosmology, West Virginia University, Chestnut Ridge Research Building, Morgantown, WV 26505, USA}
\author[0000-0001-7697-7422]{Maura A.\ McLaughlin}
\affiliation{Department of Physics and Astronomy, West Virginia University, P.O.\ Box 6315, Morgantown, WV 26506, USA}
\affiliation{Center for Gravitational Waves and Cosmology, West Virginia University, Chestnut Ridge Research Building, Morgantown, WV 26505, USA}
\author[0000-0001-5134-3925]{Gabriella Agazie}
\affiliation{Center for Gravitation, Cosmology and Astrophysics, Department of Physics, University of Wisconsin-Milwaukee,\\ P.O.\ Box 413, Milwaukee, WI 53201, USA}
\author[0000-0002-8935-9882]{Akash Anumarlapudi}
\affiliation{Center for Gravitation, Cosmology and Astrophysics, Department of Physics, University of Wisconsin-Milwaukee,\\ P.O.\ Box 413, Milwaukee, WI 53201, USA}
\author[0000-0003-0638-3340]{Anne M.\ Archibald}
\affiliation{Newcastle University, NE1 7RU, UK}
\author[0009-0008-6187-8753]{Zaven Arzoumanian}
\affiliation{X-Ray Astrophysics Laboratory, NASA Goddard Space Flight Center, Code 662, Greenbelt, MD 20771, USA}
\author[0000-0003-2745-753X]{Paul T.\ Baker}
\affiliation{Department of Physics and Astronomy, Widener University, One University Place, Chester, PA 19013, USA}
\author[0000-0003-3053-6538]{Paul R.\ Brook}
\affiliation{Institute for Gravitational Wave Astronomy and School of Physics and Astronomy, University of Birmingham, Edgbaston, Birmingham B15 2TT, UK}
\author[0000-0002-6039-692X]{H.\ Thankful Cromartie}
\affiliation{National Research Council Research Associate, National Academy of Sciences, Washington, DC 20001, USA resident at Naval Research Laboratory, Washington, DC 20375, USA}
\author[0000-0002-1529-5169]{Kathryn Crowter}
\affiliation{Department of Physics and Astronomy, University of British Columbia, 6224 Agricultural Road, Vancouver, BC V6T 1Z1, Canada}
\author[0000-0002-2185-1790]{Megan E.\ DeCesar}
\affiliation{George Mason University, Fairfax, VA 22030, resident at the U.S.\ Naval Research Laboratory, Washington, DC 20375, USA}
\author[0000-0002-6664-965X]{Paul B.\ Demorest}
\affiliation{National Radio Astronomy Observatory, 1003 Lopezville Rd., Socorro, NM 87801, USA}
\author[0000-0002-2554-0674]{Lankeswar Dey}
\affiliation{Department of Physics and Astronomy, West Virginia University, P.O.\ Box 6315, Morgantown, WV 26506, USA}
\affiliation{Center for Gravitational Waves and Cosmology, West Virginia University, Chestnut Ridge Research Building, Morgantown, WV 26505, USA}
\author[0000-0001-8885-6388]{Timothy Dolch}
\affiliation{Department of Physics, Hillsdale College, 33 E.\ College Street, Hillsdale, MI 49242, USA}
\affiliation{Eureka Scientific, 2452 Delmer Street, Suite 100, Oakland, CA 94602-3017, USA}
\author[0000-0001-7828-7708]{Elizabeth C.\ Ferrara}
\affiliation{Department of Astronomy, University of Maryland, College Park, MD 20742, USA}
\affiliation{Center for Research and Exploration in Space Science and Technology, NASA/GSFC, Greenbelt, MD 20771}
\affiliation{NASA Goddard Space Flight Center, Greenbelt, MD 20771, USA}
\author[0000-0001-8384-5049]{Emmanuel Fonseca}
\affiliation{Department of Physics and Astronomy, West Virginia University, P.O.\ Box 6315, Morgantown, WV 26506, USA}
\affiliation{Center for Gravitational Waves and Cosmology, West Virginia University, Chestnut Ridge Research Building, Morgantown, WV 26505, USA}
\author[0000-0001-7624-4616]{Gabriel E.\ Freedman}
\affiliation{Center for Gravitation, Cosmology and Astrophysics, Department of Physics, University of Wisconsin-Milwaukee,\\ P.O.\ Box 413, Milwaukee, WI 53201, USA}
\author[0000-0001-6166-9646]{Nate Garver-Daniels}
\affiliation{Department of Physics and Astronomy, West Virginia University, P.O.\ Box 6315, Morgantown, WV 26506, USA}
\affiliation{Center for Gravitational Waves and Cosmology, West Virginia University, Chestnut Ridge Research Building, Morgantown, WV 26505, USA}
\author[0000-0001-8158-683X]{Peter A.\ Gentile}
\affiliation{Department of Physics and Astronomy, West Virginia University, P.O.\ Box 6315, Morgantown, WV 26506, USA}
\affiliation{Center for Gravitational Waves and Cosmology, West Virginia University, Chestnut Ridge Research Building, Morgantown, WV 26505, USA}
\author[0000-0003-4090-9780]{Joseph Glaser}
\affiliation{Department of Physics and Astronomy, West Virginia University, P.O.\ Box 6315, Morgantown, WV 26506, USA}
\affiliation{Center for Gravitational Waves and Cosmology, West Virginia University, Chestnut Ridge Research Building, Morgantown, WV 26505, USA}
\author[0000-0003-1884-348X]{Deborah C.\ Good}
\affiliation{Department of Physics and Astronomy, University of Montana, 32 Campus Drive, Missoula, MT 59812}
\author[0000-0003-2742-3321]{Jeffrey S.\ Hazboun}
\affiliation{Department of Physics, Oregon State University, Corvallis, OR 97331, USA}
\author[0000-0003-1082-2342]{Ross J.\ Jennings}
\altaffiliation{NANOGrav Physics Frontiers Center Postdoctoral Fellow}
\affiliation{Department of Physics and Astronomy, West Virginia University, P.O.\ Box 6315, Morgantown, WV 26506, USA}
\affiliation{Center for Gravitational Waves and Cosmology, West Virginia University, Chestnut Ridge Research Building, Morgantown, WV 26505, USA}
\author[0000-0001-6607-3710]{Megan L.\ Jones}
\affiliation{Center for Gravitation, Cosmology and Astrophysics, Department of Physics, University of Wisconsin-Milwaukee,\\ P.O.\ Box 413, Milwaukee, WI 53201, USA}
\author[0000-0001-6295-2881]{David L.\ Kaplan}
\affiliation{Center for Gravitation, Cosmology and Astrophysics, Department of Physics, University of Wisconsin-Milwaukee,\\ P.O.\ Box 413, Milwaukee, WI 53201, USA}
\author[0000-0002-0893-4073]{Matthew Kerr}
\affiliation{Space Science Division, Naval Research Laboratory, Washington, DC 20375-5352, USA}
\author[0000-0003-0721-651X]{Michael T.\ Lam}
\affiliation{SETI Institute, 339 N Bernardo Ave Suite 200, Mountain View, CA 94043, USA}
\affiliation{School of Physics and Astronomy, Rochester Institute of Technology, Rochester, NY 14623, USA}
\affiliation{Laboratory for Multiwavelength Astrophysics, Rochester Institute of Technology, Rochester, NY 14623, USA}
\author[0000-0003-1301-966X]{Duncan R.\ Lorimer}
\affiliation{Department of Physics and Astronomy, West Virginia University, P.O.\ Box 6315, Morgantown, WV 26506, USA}
\affiliation{Center for Gravitational Waves and Cosmology, West Virginia University, Chestnut Ridge Research Building, Morgantown, WV 26505, USA}
\author[0000-0001-5373-5914]{Jing Luo}
\altaffiliation{Deceased}
\affiliation{Department of Astronomy \& Astrophysics, University of Toronto, 50 Saint George Street, Toronto, ON M5S 3H4, Canada}
\author[0000-0001-5229-7430]{Ryan S.\ Lynch}
\affiliation{Green Bank Observatory, P.O.\ Box 2, Green Bank, WV 24944, USA}
\author[0000-0001-5481-7559]{Alexander McEwen}
\affiliation{Center for Gravitation, Cosmology and Astrophysics, Department of Physics, University of Wisconsin-Milwaukee,\\ P.O.\ Box 413, Milwaukee, WI 53201, USA}
\author[0000-0002-4642-1260]{Natasha McMann}
\affiliation{Department of Physics and Astronomy, Vanderbilt University, 2301 Vanderbilt Place, Nashville, TN 37235, USA}
\author[0000-0001-8845-1225]{Bradley W.\ Meyers}
\affiliation{Department of Physics and Astronomy, University of British Columbia, 6224 Agricultural Road, Vancouver, BC V6T 1Z1, Canada}
\affiliation{International Centre for Radio Astronomy Research, Curtin University, Bentley, WA 6102, Australia}
\author[0000-0002-3616-5160]{Cherry Ng}
\affiliation{Dunlap Institute for Astronomy and Astrophysics, University of Toronto, 50 St.\ George St., Toronto, ON M5S 3H4, Canada}
\author[0000-0002-6709-2566]{David J.\ Nice}
\affiliation{Department of Physics, Lafayette College, Easton, PA 18042, USA}
\author[0000-0001-5465-2889]{Timothy T.\ Pennucci}
\affiliation{Institute of Physics and Astronomy, E\"{o}tv\"{o}s Lor\'{a}nd University, P\'{a}zm\'{a}ny P.\ s.\ 1/A, 1117 Budapest, Hungary}
\author[0000-0002-8509-5947]{Benetge B.\ P.\ Perera}
\affiliation{Arecibo Observatory, HC3 Box 53995, Arecibo, PR 00612, USA}
\author[0000-0002-8826-1285]{Nihan S.\ Pol}
\affiliation{Department of Physics and Astronomy, Vanderbilt University, 2301 Vanderbilt Place, Nashville, TN 37235, USA}
\author[0000-0002-2074-4360]{Henri A.\ Radovan}
\affiliation{Department of Physics, University of Puerto Rico, Mayag\"{u}ez, PR 00681, USA}
\author[0000-0001-5799-9714]{Scott M.\ Ransom}
\affiliation{National Radio Astronomy Observatory, 520 Edgemont Road, Charlottesville, VA 22903, USA}
\author[0000-0002-5297-5278]{Paul S.\ Ray}
\affiliation{Space Science Division, Naval Research Laboratory, Washington, DC 20375-5352, USA}
\author[0000-0003-4391-936X]{Ann Schmiedekamp}
\affiliation{Department of Physics, Penn State Abington, Abington, PA 19001, USA}
\author[0000-0002-1283-2184]{Carl Schmiedekamp}
\affiliation{Department of Physics, Penn State Abington, Abington, PA 19001, USA}
\author[0000-0002-7283-1124]{Brent J.\ Shapiro-Albert}
\affiliation{Department of Physics and Astronomy, West Virginia University, P.O.\ Box 6315, Morgantown, WV 26506, USA}
\affiliation{Center for Gravitational Waves and Cosmology, West Virginia University, Chestnut Ridge Research Building, Morgantown, WV 26505, USA}
\affiliation{Giant Army, 915A 17th Ave, Seattle WA 98122}
\author[0000-0001-9784-8670]{Ingrid H.\ Stairs}
\affiliation{Department of Physics and Astronomy, University of British Columbia, 6224 Agricultural Road, Vancouver, BC V6T 1Z1, Canada}
\author[0000-0002-7261-594X]{Kevin Stovall}
\affiliation{National Radio Astronomy Observatory, 1003 Lopezville Rd., Socorro, NM 87801, USA}
\author[0000-0002-2820-0931]{Abhimanyu Susobhanan}
\affiliation{Max-Planck-Institut f\"{u}r Gravitationsphysik (Albert-Einstein-Institut), Callinstrasse 38, D-30167, Hannover, Germany}
\author[0000-0002-1075-3837]{Joseph K.\ Swiggum}
\altaffiliation{NANOGrav Physics Frontiers Center Postdoctoral Fellow}
\affiliation{Department of Physics, Lafayette College, Easton, PA 18042, USA}
\author[0000-0001-9678-0299]{Haley M.\ Wahl}
\affiliation{Department of Physics and Astronomy, West Virginia University, P.O.\ Box 6315, Morgantown, WV 26506, USA}
\affiliation{Center for Gravitational Waves and Cosmology, West Virginia University, Chestnut Ridge Research Building, Morgantown, WV 26505, USA}

\begin{abstract}

Pulse profile stability is a central assumption of standard pulsar timing methods. Thus, it is important for pulsar timing array experiments such as the North American Nanohertz Observatory for Gravitational Waves (NANOGrav) to account for any pulse profile variability present in their data sets. We show that in the NANOGrav 15-yr data set, the integrated pulse profile of PSR J1022+1001 as seen by the Arecibo radio telescope at 430, 1380, and 2030 MHz varies considerably in its shape from observation to observation. We investigate the possibility that this is due to the ``ideal feed assumption'' (IFA), on which NANOGrav's routine polarization calibration procedure relies. PSR J1022+1001 is $\sim 90\%$ polarized in one pulse profile component, and also has significant levels of circular polarization. Time-dependent deviations in the feed's polarimetric response (PR) could cause mixing between the intensity $I$ and the other Stokes parameters, leading to the observed variability. We calibrate the PR using a mixture of Measurement Equation Modeling and Measurement Equation Template Matching techniques. The resulting profiles are no less variable than those calibrated using the IFA method, nor do they provide an improvement in the timing quality of this pulsar. We observe the pulse shape in 25-MHz bandwidths to vary consistently across the band, which cannot be explained by interstellar scintillation in combination with profile evolution with frequency. Instead, we favor phenomena intrinsic to the pulsar as the cause.

\end{abstract}

\section{Introduction} \label{sec:intro}

In 2023, pulsar timing array \citep[PTA;][]{fb+90} collaborations around the world released data sets presenting evidence of a stochastic background of gravitational waves (GWs) at $\sim$nanohertz frequencies \citep{NG15gwb,EPTADR2gwb,PPTADR3gwb,xcg+23}. For the past several years, these ongoing experiments have regularly observed dozens of radio pulsars in our Galaxy, in an effort to detect and characterize the properties of this GW background (GWB), determine the source(s) of the GWB, and detect continuous GWs from inspiraling supermassive black hole binaries. The North American Nanohertz Observatory for Gravitational Waves \citep[NANOGrav;][]{sbc+19} is one such collaboration, and its most recent data release is the 15-yr data set \citep[][hereafter NG15]{NG15data}.

The regular rotations of pulsars, when paired with regular ($\sim$monthly) observations over a span of months to years, allow us to unambiguously account for each rotation of the neutron star.
This is achieved by constructing a \textit{timing model}, which is composed of any and all parameters that measurably impact the times of arrival (TOAs) of the pulsar signal at the telescope. This includes (but is not limited to) rotational, astrometric, and binary (if applicable) parameters.
The TOAs predicted by the model are subtracted from those measured by observation to obtain timing \textit{residuals}, which are used to refine the timing model.

A GW signal arises in PTA data as a red noise process in the timing residuals, and should be correlated between the pulsars in a characteristic way \citep{NG15gwb}. It is, then, critically important to characterize all sources of noise and systematic error in pulsar timing residuals, so as not to confuse noise of non-GW origin for a detection, and maintain the best sensitivity to GWs \citep{NG15noise}. Thus, it is important to investigate any unusual behavior by a pulsar in a PTA data set that could potentially give rise to unaccounted-for noise.

\subsection{PSR J1022+1001}

\begin{figure}
    \centering
    \includegraphics[width=1\linewidth]{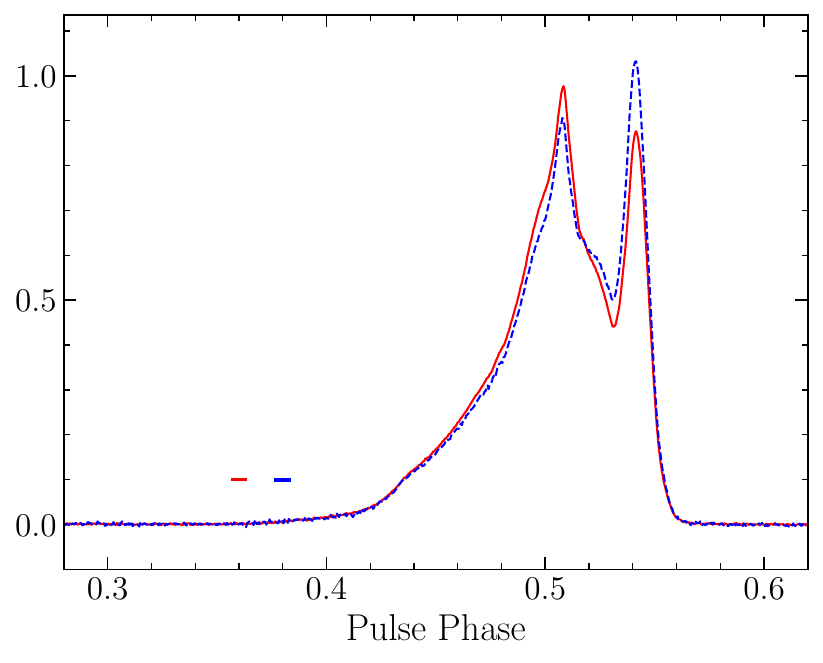}
    \caption{Two example pulse profiles of PSR J1022+1001 from 1.4-GHz Arecibo observations in the NANOGrav 15-yr data set. The solid red profile corresponds to an observation made on 5 November 2015, while the dashed blue profile was obtained on 5 April 2020; these observations were chosen to illustrate the stark difference in peak heights between the two pulse profiles. The profiles were normalized using the area under the curve. The pulse phase is given in units of rotations. The small red and blue ``lines'' in the bottom left of the figure are actually error bars showing the level of off-pulse noise in each profile.}
    \label{fig:example}
\end{figure}

PSR J1022+1001, is an MSP with $P = 16.45\,\ms$ and a dispersion measure (DM) of 10.26\,\dmu, in a 7.8\,day orbit with a white dwarf companion \citep{cns+96}. It is observed by the European Pulsar Timing Array \citep[EPTA;][]{EPTADR2data}, Parkes Pulsar Timing Array \citep[PPTA;][]{PPTADR3data}, and NANOGrav. Since shortly after its discovery, PSR J1022+1001 has been known to exhibit strange pulse shape variations from observation to observation \citep{cns+96,kxc+99}. This is illustrated in Figure~\ref{fig:example}, which shows integrated pulse profiles obtained from two NANOGrav observations of PSR J1022+1001 with the 305-m radio telescope at the Arecibo Observatory (AO), one of which was performed 4.4\,yr after the other. A clear difference in the pulse shape can be seen, such that the first peak is higher than the second in one case, while the reverse is true in the other. This variability has been studied over nearly three decades, with contradictory conclusions as to the underlying cause, or whether there is even variability at all \citep{kxc+99,rk+03,hbo+04,vsw+13,lkl+15,sy+16,pbc+21}. 

When pulsar data are folded over several thousand rotations, the resulting pulse profile is generally long-term stable, as the fluctuations in the shapes of individual pulses average out \citep[this stability is, however, limited by pulse jitter;][]{lma+19}. However, integrated pulse profiles have been observed to vary in some pulsars due to various causes. Some pulsars exhibit forms of pulse shape variability such as \textit{mode changing}, where on timescales of several to thousands of rotations, the pulse profile changes between two or more distinct shapes \citep{bdc+70,wmj+07}. Pulse profiles have also been seen to change over time for binary pulsars due to geodetic precession \citep[e.g.][]{cbc+23}. 

In the case of PSR J1713+0747, a sudden change was observed in the pulse profile, which then gradually approached its previous shape over the course of several months \citep{jcc+24}. Geodetic precession was ruled out as a cause, but an extreme case of mode changing was not. Other proposed causes include plasma lensing in the ISM (not ruled out), a shape change associated with a glitch (ruled out), and more exotic phenomena such as the incursion of an asteroid into the pulsar magnetosphere. While there is no single satisfying explanation, some event related to the pulsar magnetosphere seems most likely \citep{jcc+24}.

The profile variability seen for PSR J1022+1001 does not fit neatly into any of the above categories. It is seemingly random, without any distinct modes or long-term trends \citep{kxc+99}. Notably, PSR J1022+1001's pulse profile is highly polarized, with the polarization fraction reaching $\sim$90\% in the second peak of emission. Thus, one hypothesized cause of the observed phenomenon is mixing between polarized and unpolarized intensities caused by polarization mis-calibration \citep{kxc+99,vsw+13}. The main purpose of this paper is to use NANOGrav timing observations to evaluate the likelihood of this possibility.

\subsection{Pulsar Polarimetry}

The intensity of a polarized electromagnetic wave is described by the Stokes vector

\begin{equation}
  \mathbf{S} = \begin{bmatrix}
  I \\
  Q \\
  U \\
  V \\
\end{bmatrix}\;,
\end{equation}
where $I$ is the total intensity, $Q$ and $U$ represent two orthogonal components of linearly-polarized emission, and $V$ is the circularly-polarized intensity. The linearly-polarized intensity is given by $L = \sqrt{Q^2 + U^2}$, and the polarization angle ($\Psi$ or P.A.) is given by \begin{equation} \label{eq:PA}
    \Psi = 0.5\,\tan^{-1}\left({\frac{U}{Q}}\right)\,.
\end{equation}

In this work, we employ the IAU convention for P.A.\ and $V$, where $\Psi = 0\arcdeg$ points north, P.A.\ increases in the counter-clockwise direction, and $V > 0$ represents a rotation that increases the P.A.\ (a right-handed circularly polarized signal).

In the context of pulsars, the dependence of $\Psi$ on pulse longitude $\varphi$ is used to glean insights into the emission geometry. Especially in slow pulsars, $\Psi$ often follows an ``S''-shaped curve. This is explained by the Rotating Vector Model \citep[RVM;][]{rc+69}, which has $\Psi$ determined by the projection of the pulsar's magnetic field lines along the line of sight:
\begin{equation} \label{eq:RVM}
    \tan(\Psi - \Psi_0) = \frac{\sin\alpha\sin(\varphi - \varphi_0)}{\sin\zeta\cos\alpha - \cos\zeta\sin\alpha\cos(\varphi-\varphi_0)},
\end{equation}
where $\Psi_0$ and $\varphi_0$ are values of $\Psi$ and $\varphi$ where the gradient of $\Psi$ is greatest, defining the point of symmetry for the curve such that the pulsar beam is pointing most directly towards the observer; $\alpha$ is the angle between the pulsar's axis of rotation and the magnetic pole; and $\zeta$ is the angle between the rotation axis and the line of sight \citep{ew+01}. Therefore, the minimum angle between the magnetic axis and the line of sight is
\begin{equation} \label{eq:beta}
    \beta = \zeta - \alpha\,.
\end{equation}

The P.A.\ recorded by a telescope on Earth must be corrected for Faraday rotation to obtain the true P.A.\ of the emitted signal. 
The component of the magnetic field along the line of sight to the pulsar, $B_{||}$, subjects the emission's P.A.\ to a wavelength-dependent rotation:
\begin{equation}
    \Psi(\lambda) - \Psi_\infty = \RM~\lambda^2\,,
\end{equation}
where $\Psi_\infty$ is the P.A.\ at infinite frequency, $\lambda$ is the wavelength of the radio waves and $\RM$ is the rotation measure, which is given (in Gaussian cgs units) by
\begin{equation}
    \RM = \frac{e^3}{2\pi m_\textrm{e}^2 c^4} \int_{0}^{d} n_\textrm{e}(l) B_{||}(l)\,dl%\,,
\end{equation}
where $n_\textrm{e}$ is the density of free electrons, $e$ and $m_\textrm{e}$ are the charge and mass of the electron, respectively, $c$ is the speed of light, and $d$ is the pulsar distance \citep{lk+04}. 

After correcting for Faraday rotation, the Stokes vector emitted by a polarized radio source ($\mathbf{S}_\textrm{emitted}$) differs from that which is recorded by a radio telescope ($\mathbf{S}_\textrm{meas}$) by a linear transformation:
\begin{equation} \label{eq:mueller}
    \mathbf{S}_\textrm{meas} = \mathbf{M}~\mathbf{S}_\textrm{emitted}\,.
\end{equation}
Here, the Mueller matrix $\mathbf{M}$ is effectively the transfer function for $\mathbf{S}$ \citep{hpn+01}, also known as the polarimetric response (PR) of the feed. 
We parameterize the PR as in Equation 18 of \citet{vsw+04}, with the absolute gain $G$, differential gain $\gamma$, differential phase $\phi$, and the orientations $\theta_k$ and ellipticities $\epsilon_k$ of receptors $k \in 0, 1$.

These parameters, which correspond to physical qualities of the feed, describe mixing between the components of $\mathbf{S}$. For example, assuming a linear feed, $\gamma$ describes mixing between $I$ and $Q$, while $\phi$ describes rotation about the $Q$ axis. They are known to vary with time \citep[e.g.][]{gmd+18}, which could in turn cause $\mathbf{S}_\textrm{meas}$ to vary with time even if $\mathbf{S}_\textrm{emitted}$ is stable.

In this work, we present the pulse profile variability seen in NANOGrav observations of PSR J1022+1001, and investigate whether robust polarization calibration can account for the observed variability. Observations of PSR J1022+1001 and the pulsars used for polarization calibration purposes are described in \S\ref{sec:data}. Our methods of data reduction, polarization calibration, and pulsar timing are detailed in \S\ref{sec:methods}. We show our results in \S\ref{sec:results}. Finally, we discuss our results in context of prior work and conclude with \S\ref{sec:discussion}.

\section{Observations} \label{sec:data}

This work makes use of a subset of observations released as part of NG15, along with several observations of the slow pulsar B0525+21 (J0528+2200). Without exception, these data were taken at AO using the Puerto Rican Ultimate Pulsar Processing Instrument \citep[PUPPI][]{vws+17}, which is a clone of the Green Bank Ultimate Pulsar Processing Instrument (GUPPI), the pulsar backend formerly used by the Green Bank Telescope \citep{rdf+09}. 

Observations were made with the 430 (circular feeds; 421--445\,MHz bandwidth split into 1.5625-MHz wide frequency channels), L-wide (linear feeds; 1147--1765\,MHz; 12.5\,MHz channels) and S-wide (linear feeds; concontiguous bands at 1700--1880 and 2050--2404\,MHz; 12.5\,MHz channels) receivers.
The data were coherently folded and de-dispersed using the pulsar ephemerides, resulting in 10\,s sub-integrations and 2048 pulse phase bins. 
We generally refer to the data using the approximate center frequencies instead of the receiver names: 430\,MHz, 1.4\,GHz, and 2\,GHz data for the 430, L-wide, and S-wide recievers, respectively.

For a detailed description of NANOGrav timing observations and the 15-yr data set, see NG15. For our analysis, we used all AO observations of PSR J1022+1001 as our primary data of interest. PSR J1022+1001 was observed at AO 84 times between MJDs 57028 and 59055 (a $\approx$24-day cadence). On each observing epoch, the pulsar was observed at both 430\,MHz and 1.4\,GHz (until MJD 57502) or 1.4 and 2\,GHz (beginning MJD 57522) for about 20--25\,min with each receiver.

For polarization calibration purposes (described in \S\ref{sec:METM}), we also made use of all 430-MHz observations of PSR J0030+0451 and all 1.4-GHz and 2-GHz observations of PSR B1937+21 (J1939+2134). Table~\ref{tab:mem} provides a summary of the observations of PSR B0525+21, which were not part of the NANOGrav 15-yr data release. These observations were purely for polarization calibration purposes, which are described in \S\ref{sec:MEM}.

Data taken between 2017 August and 2018 November were affected by a malfunctioning local oscillator (LO) at AO. This issue is described in further detail by NG15. As in that work, we disregarded all observations that fell within this time range.

\begin{deluxetable}{cccccccc}
  \tabletypesize{\footnotesize}
  \tablewidth{\textwidth}
  \tablecaption{Observations of PSR B0525+21}
  \tablecolumns{8}
  % \vspace{5mm}
  \tablehead{
  \colhead{Rcvr} & 
  \colhead{$\nu$} &
  \colhead{$\Delta\nu_\textrm{usable}$} &
  \colhead{$\Delta\nu_\textrm{chan}$} &
  \colhead{MJD} &
  \colhead{$N_\textrm{obs}$} &
  \colhead{$t_\textrm{int}$} &
  \colhead{Span} \\
  \colhead{} &
  \colhead{(GHz)} &
  \colhead{(MHz)} &
  \colhead{(MHz)} &
  \colhead{} &
  \colhead{} &
  \colhead{(min)} &
  \colhead{(min)}
  }
  \startdata
  430 & 0.43 & 24 & 1.5625 & 56744 & 4 & 75.5 & 129.3 \\
  L-wide & 1.4\phn & 600 & 12.5 & 56116 & 3 & 56.2 & 109.3 \\
  S-wide & 2.0\phn & 460 & 12.5 & 56116 & 3 & 31.4 & 112.7
  \enddata
  \tablecomments{Summary of observations of PSR B0525+21 used to generate MEM receiver solutions. From left to right, columns correspond to receiver name, approximate center frequency, usable bandwidth, frequency channel bandwidth, MJD of the observations, number of observations, cumulative length of observations, and timespan between the start of the first observation and the end of the last observation. Aside from the total integration time, these observing setups are identical to those used by NANOGrav timing observations with these receivers.}
  \label{tab:mem}
\end{deluxetable}

\section{Methods} \label{sec:methods}

In \S\ref{sec:polcal}, we describe the methods used to produce polarization-calibrated profiles. In \S\ref{sec:timing}, we describe our data reduction procedure and timing analysis of the calibrated data.

\subsection{Polarization Calibration} \label{sec:polcal}

In order to calibrate our observations of PSR J1022+1001, we used a combination of Measurement Equation Modeling \citep[MEM;][]{vsw+04} and Measurement Equation Template Matching \citep[METM;][]{vsw+13} methods. For a given receiver, we applied MEM to observations of the slow pulsar B0525+21 to produce a PR. Then, METM was used to calculate ``corrections'' to this PR using observations of ``template pulsars'' made on different epochs. The MEM PR, along with METM corrections, were then used to calibrate the observations of PSR J1022+1001.

In \S\ref{sec:IFA}, we first describe the usual method of polarization calibration employed in NANOGrav data reduction, which is known as the ``Ideal Feed Assumption,'' or IFA. Then, we describe the process of generating receiver solutions using MEM (\S\ref{sec:MEM}) and METM (\S\ref{sec:METM}). In each case, the data were calibrated using \psrchive\footnote{\url{http://psrchive.sourceforge.net/}} \citep{psrchive}, in particular \texttt{pac}. In the cases of MEM and METM, \texttt{pcm} was first used in concert with calibration observations to generate PRs.

\subsubsection{Ideal Feed Assumption} \label{sec:IFA}

The standard method of polarization calibration for NANOGrav timing observations aims to account for differences in gain and phase between the two receptors by observing a reference signal of artificial origin. A noise diode coupled to the receiver injects a polarized square wave signal of known amplitude and position angle, which is used to determine the absolute gain $G$, differential gain $\gamma$, and differential phase $\phi$ of the receptors. Approximately once per month, flux calibration is performed by measuring the noise diode signal simultaneously with on/off observations of a quasar (J1445+0958) whose emission is unpolarized and of a known flux density, which calibrates the amplitude of the reference signal and refines the determination of $G$.

This method assumes the receptors are perfectly orthogonal, the noise diode signal is 100\% linearly polarized, and that it illuminates both receptors equally and in phase. However, flux calibration removes the need to assume the signal illuminates the receptors equally \citep{vsw+02}.

If the receptors are not perfectly orthogonal, the resulting cross-coupling between the two polarizations causes mixing between total intensity and polarized signal. For a highly-polarized pulsar signal, this could lead to a temporal variation in the total intensity pulse profile, as is seen for PSR J1022+1001 \citep{vsw+13}.

\subsubsection{Measurement Equation Modeling} \label{sec:MEM}

MEM uses one or multiple observations of a suitably-polarized pulsar that span a range of parallactic angles, along with an artificial noise diode signal (as with IFA), to fully characterize the PR of the receiver without requiring the assumptions made in IFA calibration. A first guess of the PR is obtained using the noise diode observation. A set of pulse phase bins are then chosen to model the dependency of the Stokes parameters on parallactic angle. These curves are then fitted to the polarization measurement equation \citep{hjp+00}, and the best-fit solution gives the PR in the chosen parameterization. A flux calibrator observation (as described in Section~\ref{sec:IFA}) is used to break measurement equation degeneracy by constraining the mixing of Stokes $I$ and either $Q$ (for a linear feed) or $V$ (for a circular feed). MEM is described in complete detail in \citet{vsw+04}.

\begin{figure*}
    \centering
    \includegraphics[width=0.9\textwidth]{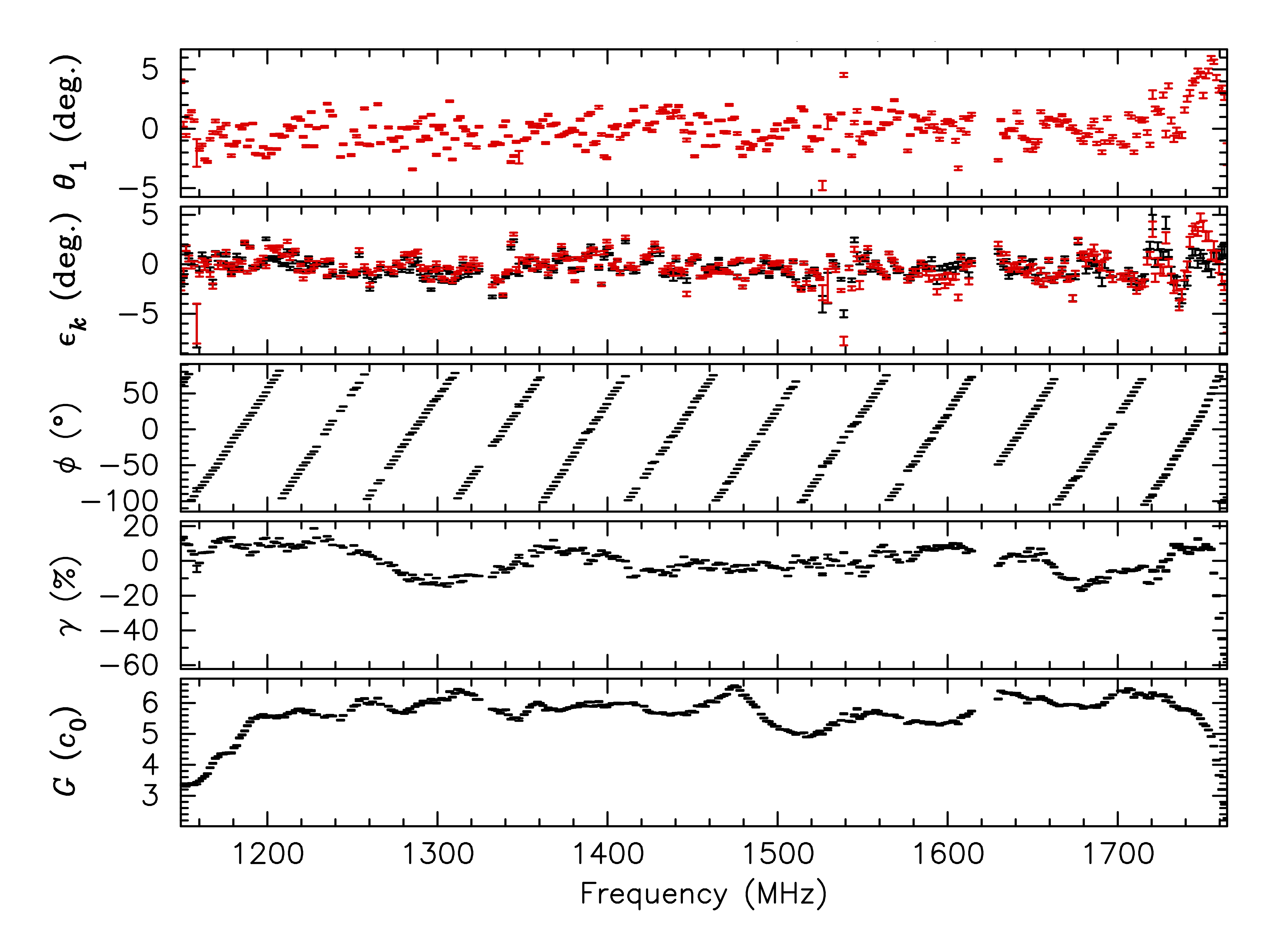}
    \caption{Polarimetric response of the L-wide (1.4\,GHz) receiver on 7 August 2012 (MJD 56116). The parameters shown are $\theta_1$, the orientation of receptor 1 with respect to receptor 0; $\epsilon_k$, the ellipticities of the receptors (receptor 0 in black and receptor 1 in red); $\phi$, the differential phase; $\gamma$, the differential gain; and $G$, the absolute gain.}
    \label{fig:MEM}
\end{figure*}

One MEM receiver solution was obtained for each receiver using observations of the slow pulsar B0525+21. These observations are described in Table~\ref{tab:mem}. The resulting PR for the L-wide receiver (for MJD 56116) is shown in Figure~\ref{fig:MEM}.

\subsubsection{Measurement Equation Template Matching} \label{sec:METM}

METM calibration compares observations of a pulsar to a single calibrated ``template'' profile of the same pulsar, obtaining PRs on multiple epochs. Assuming the profile of this pulsar does not change with time, the PR is calculated by matching the profile in question to the template profile. This process is described fully in \citet{vsw+13}.

\begin{figure*}
    \centering
    \includegraphics[width=0.9\textwidth]{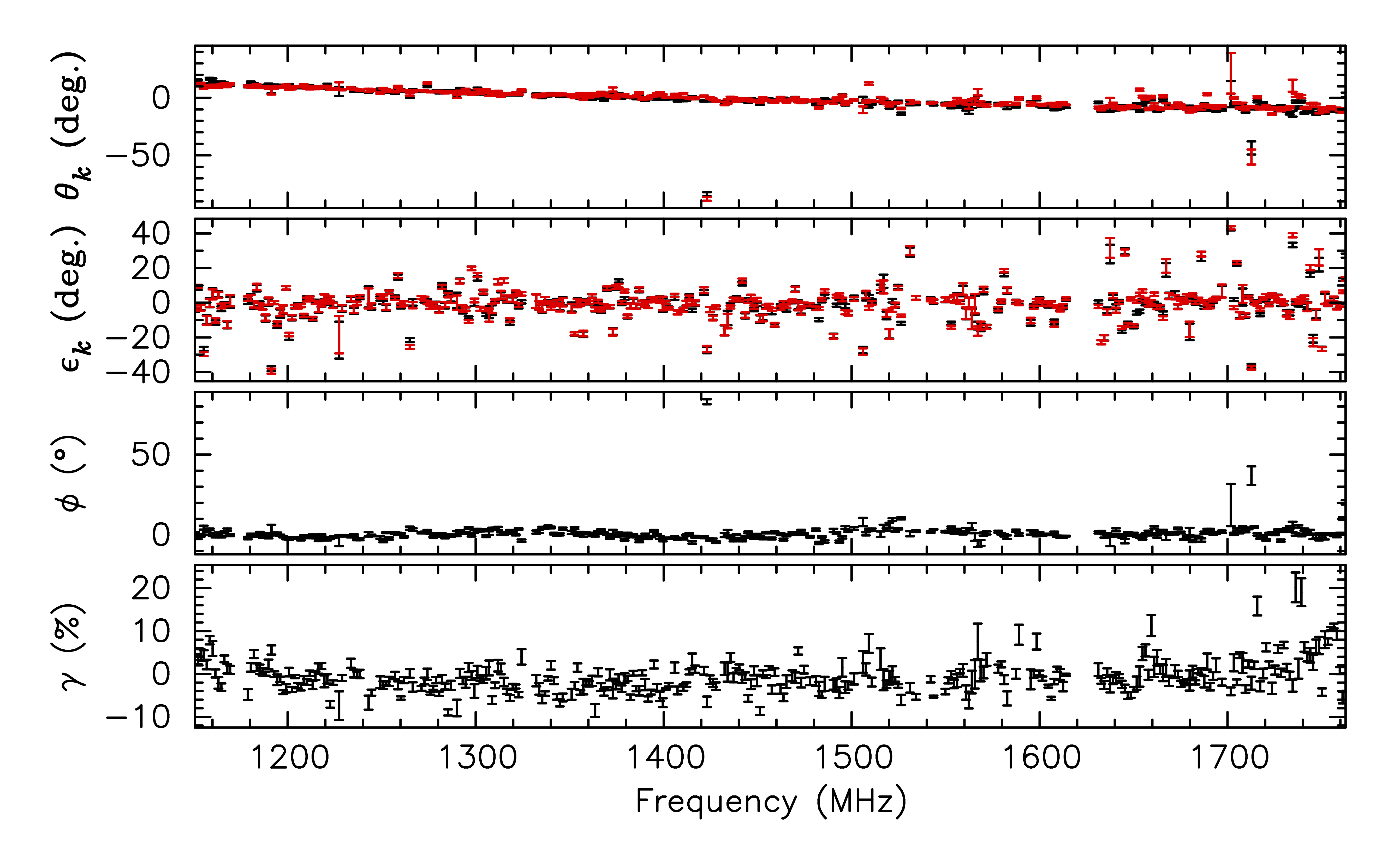}
    \caption{METM correction to the 1.4-GHz MEM PR calculated from a 31 July 2012 (MJD 56139) observation of PSR B1937+21. The parameters shown are the orientations $\theta_k$ and ellipticities $\epsilon_k$ of the receptors (receptor 0 in black and receptor 1 in red); $\phi$, the differential phase; and $\gamma$, the differential gain. Because the Stokes parameters were normalized by the mean invariant, our METM solutions do not provide any information about the absolute gain $G$.}
    \label{fig:METM}
\end{figure*}

We used PSRs J0030+0451 (at 430\,MHz) and B1937+21 (at 1.4 and 2\,GHz) as our template pulsars. These pulsars were chosen for their high S/N and significant levels of linear and circular polarization. Following \citet{gmd+18}, we apply the MEM PR to each observation of these pulsars. For each receiver, we first considered the two MEM-calibrated profiles taken closest in time to the MEM PR. In each case, these were visually indistinguishable from previously-published calibrated profiles in \citet{gmd+18} and \citet{wmg+22}. From those two profiles, we chose the profile with the highest S/N to be our template profile. These were then used as described above, to create PRs corresponding to each other template pulsar observation. These PRs can be thought of as corrections to the MEM PR, and were applied to the data of interest after first applying the MEM PR. An example METM PR is shown in Figure~\ref{fig:METM}.

Interstellar scintillation induces a frequency-dependent flux density modulation on a timescale specific to each pulsar. When comparing observations of our template pulsars to the relevant template profile, differences in the flux density in each channel due to scintillation are covariant with changes in $G$. To avoid this, we normalized the Stokes parameters by the invariant interval $I^2 - Q^2 - U^2 - V^2$ \citep{bmc+00}, trusting that $G$ is sufficiently characterized by our flux calibrator observations.

\subsection{Data Reduction \& Timing Analysis} \label{sec:timing}

We largely followed the standard NANOGrav data reduction and TOA generation procedure, as described in NG15. First, data affected by the LO issue described in \S\ref{sec:data} were excluded. Artifacts introduced by the interleaved analog-to-digital converter scheme used by the PUPPI receiver \citep{abb+21} were removed from the raw data. Then, we performed standard radio frequency interference (RFI) excision and polarization-calibrated the data as described above, followed by additional RFI excision. We performed the additional step of correcting the calibrated profiles for interstellar (and ionospheric) Faraday rotation using \texttt{rmfit}, assigning an RM to each observation. This was followed by time-averaging to maximum sub-integrations of 30\,min, and (for 1.4 and 2\,GHz data) frequency-averaging, achieving 12.5\,MHz-wide subbands (430-MHz data kept the 1.5625\,MHz-wide subbands of the raw data).

Following this, a standard template profile was created for each receiver by iteratively fitting a number of Gaussian curves to the S/N-weighted sum of the individual profiles. This Gaussian-based standard profile was then denoised with UD8 wavelet smoothing. TOAs were generated with \texttt{pat}, with a separate TOA being generated for each sub-integration and frequency subband by measuring the time shift between the profile and the template \citep{tjh+92}.

We then performed an outlier analysis on the TOAs, using the automated pipeline described in Section 3.3 of NG15. TOAs were removed if the pipeline determined an outlier probability greater than 0.1, or if the profile S/N was less than 8. Observations with $>$8\% of TOAs flagged as outliers were removed entirely. Observations where the ratio of maximum to minimum TOA frequency did not exceed 1.1 were also eliminated.

After outlier excision, we used \texttt{PINT\_Pal},\footnote{\url{https://github.com/nanograv/pint_pal}} the timing analysis package that was employed in NG15 and makes use of the pulsar timing package \pint\footnote{\url{https://github.com/nanograv/PINT}} \citep{pint+21,pint+24}, to produce a timing solution. The timing parameters and TOAs are contained within a \texttt{par} and \texttt{tim} file, respectively. \pint, and indeed all pulsar timing methods, subtracts measured TOAs from the TOAs predicted by the pulsar's timing model to obtain the timing residuals. A least-squares fit is performed on the residuals to estimate the timing model parameters. The \texttt{PINT\_Pal} timing notebooks also allow for modeling the noise characteristics of the timing residuals, and perform statistical tests to determine whether any parameters should be added to or removed from the model.

Our timing analysis closely follows NG15 \S4, which we summarize here. We model the pulsar's rotation with spin frequency, phase, and frequency derivative. Five parameters account for the astrometric position and motion of the pulsar: ecliptic longitude and latitude, proper motion in each coordinate, and timing parallax. To model timing variations caused by the pulsar's orbit with its binary companion, we used the DD binary model \citep{dd+85}. The five Keplerian parameters (orbital period, projected semimajor axis, eccentricity, and epoch and longitude of periastron) are always fit, along with the corresponding derivatives if significant. The above parameters are referenced to the midpoint of the data timespan.

Due to the turbulent ionized ISM and ionospheric and solar-wind variations, pulsar dispersion measures are time-dependent \citep{jml+17}. To account for these variations, we employ the DMX model of time-dependent dispersion measure in \pint, which is a piecewise constant function. TOAs are divided into distinct epochs of a specified length, and each epoch is assigned a single DMX parameter. Each DMX parameter describes the offset between the measured DM within an epoch and a fixed reference DM. 

Because of its low ecliptic latitude, our line of sight to PSR J1022+1001 annually intercepts the Sun. When the pulsar's angular distance from the Sun is small, the solar wind induces timing variations exceeding 100\,ns at all of our observing frequencies. Due to this, and the fact that observations were made at multiple frequencies on the same day at AO (enabling a precise measurement of DM from a single day's observations), we follow \citet{NG15data} in using DMX epochs of half-day length.

PSR J1022+1001's pulse profile is noticeably frequency-dependent within the bandwidths of the receivers, especially that of the L-wide receiver. Differences between the shape of the standard profile and the profiles corresponding to the subbanded TOAs can induce frequency-dependent timing variations. In the timing analysis, we account for the effects of these variations using frequency-dependent (FD) parameters $c_i$ (also \texttt{FD1}, \texttt{FD2}, etc.) that describe a timing delay 
\begin{equation}
    \Delta_\mathrm{FD} = \sum_{i=1}^n c_i\log\left(\frac{\nu}{1\,\GHz}\right)^i
\end{equation}
\citep{abc+15}.

Lastly, constant phase offsets between the different AO receivers were accounted for by fitting two \texttt{JUMP} parameters. We determined whether to add new timing model parameters or remove existing ones using a statistical $F$-test, including parameters only if 3-$\sigma$ confidence was reached. For consistency with NANOGrav's timing process, proper motion and parallax parameters were included regardless of significance (NG15).

When we began our timing analysis, we noted that the DM reported in the 15-yr \texttt{par} file (9.435\,\dmu) was significantly different from the value used for coherent dedispersion (10.26\,\dmu). From visual inspection of the data, it became clear that the 15-yr value was incorrect. We see similar discrepancies for other pulsars, such as a 0.08\,\dmu excess for PSR B1937+21. 
Further investigation showed this was due to a covariance between the DMX and FD parameters that led to the FD parameters absorbing a $\nu^{-2}$ delay, and biasing the DMX parameters.
This bias was absorbed into the reference DM as part of NANOGrav's iterative timing process, which sets the DMX parameters to have a mean value of zero with each iteration.

Efforts to avoid this problem in the future are underway within NANOGrav. 
In this analysis, we only added an FD parameter if doing so did not cause a large change ($\gtrsim 0.1$\,\dmu) in the DMX average, which we interpret as spurious. We did not see the same phenomenon while performing our timing analysis on the calibrated profiles.

We also performed the outlier and timing analysis steps on the observations of PSR J1022+1001 as presented in NG15, which were only subjected to the standard IFA-based polarization calibration. This amounted to redoing the same analysis as in NG15, except we avoided adding FD parameters when this resulted in a significant departure from the known DM value, even if such an addition was preferred by F-tests. Hereafter, we refer our two data sets as the IFA and METM data sets.

\section{Results} \label{sec:results}

In \S\ref{sec:variability}, we describe our analysis of profile variability in the IFA and METM data sets. In \S\ref{sec:timingsolution}, we compare the timing solutions obtained from the IFA and METM data sets. 

\subsection{Polarization Profiles} \label{sec:polarimetry}

\begin{figure}
    \centering
    \includegraphics[width=1\linewidth]{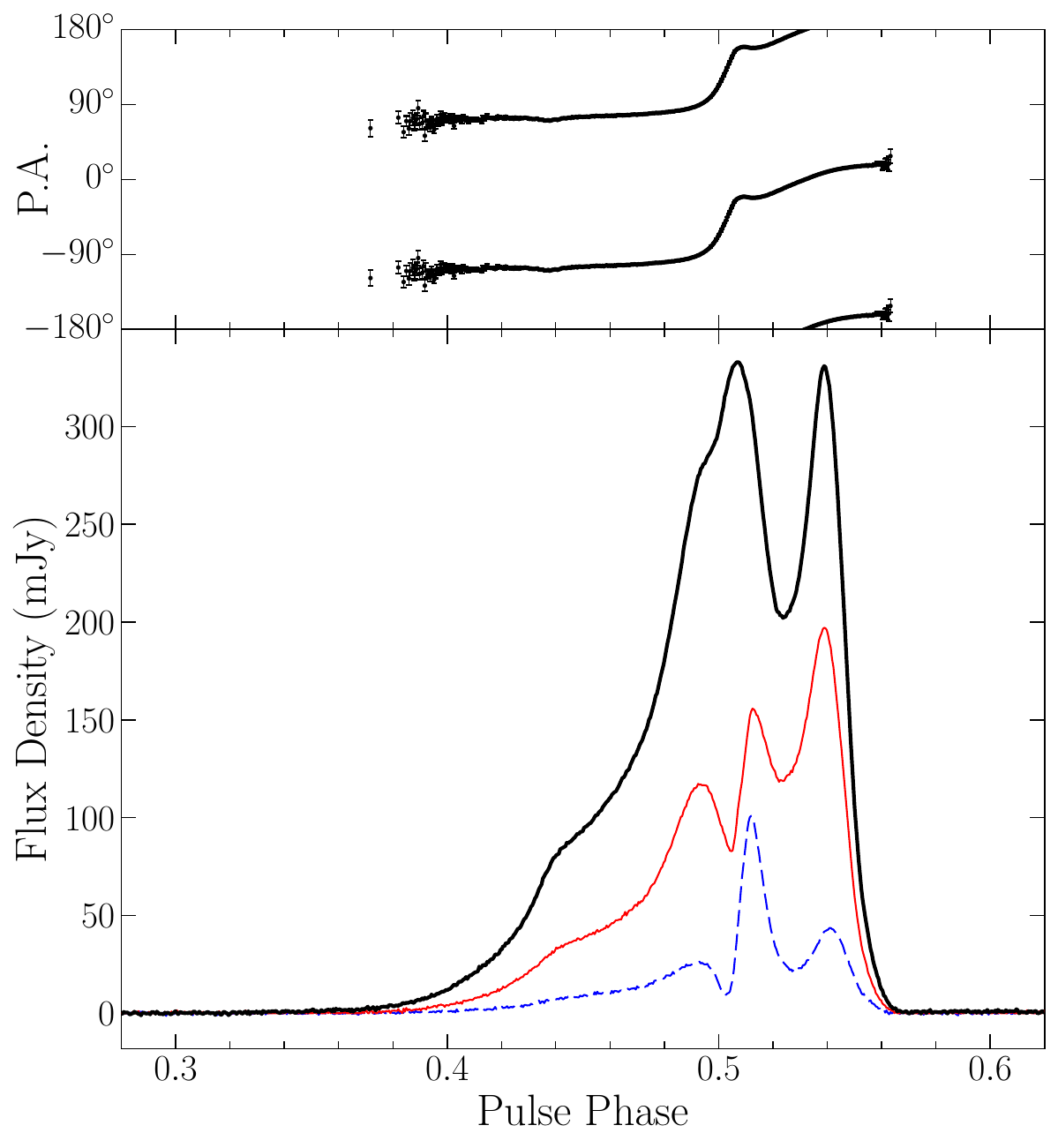}
    \caption{METM-calibrated polarization profile of PSR J1022+1001 at 430\,MHz. The bottom panel shows the total intensity profile (Stokes $I$) with the thick black line, the linear polarization ($L$) with the thin red line, and the circular polarization ($V$) with the blue dashed line. The top panel shows the P.A.\ swing in bins for which $L > 3\sigma_I$. We used the IAU convention for P.A.\ and $V$.}
    \label{fig:polprof430}
\end{figure}

\begin{figure}
    \centering
    \includegraphics[width=1\linewidth]{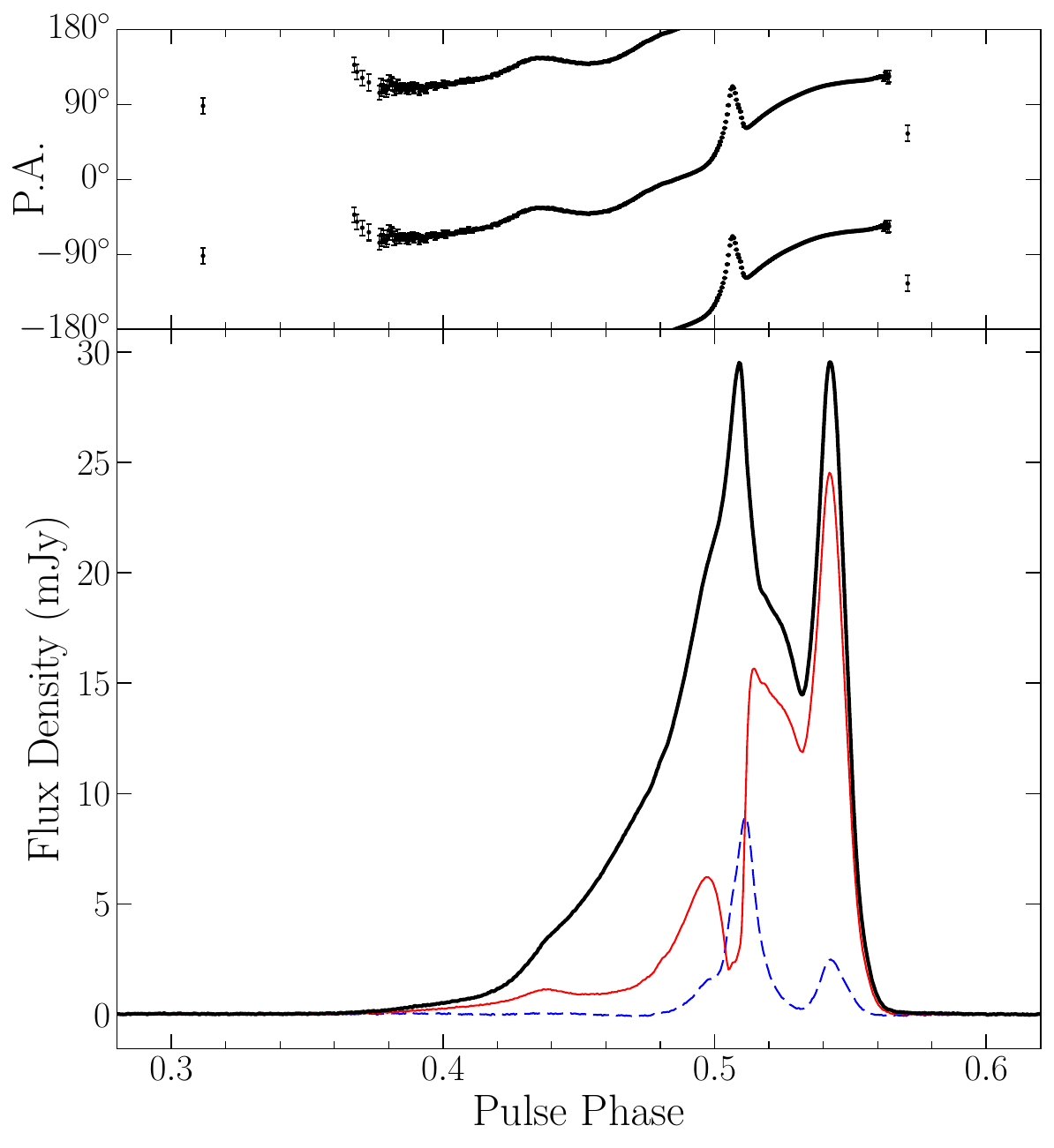}
    \caption{METM-calibrated polarization profile of PSR J1022+1001 at 1.4\,GHz. See Figure~\ref{fig:polprof430}'s caption for details. A Rotating Vector Model fit to the P.A.\ curve is also shown, and is described in the text.}
    \label{fig:polprofL}
\end{figure}

\begin{figure}
    \centering
    \includegraphics[width=1\linewidth]{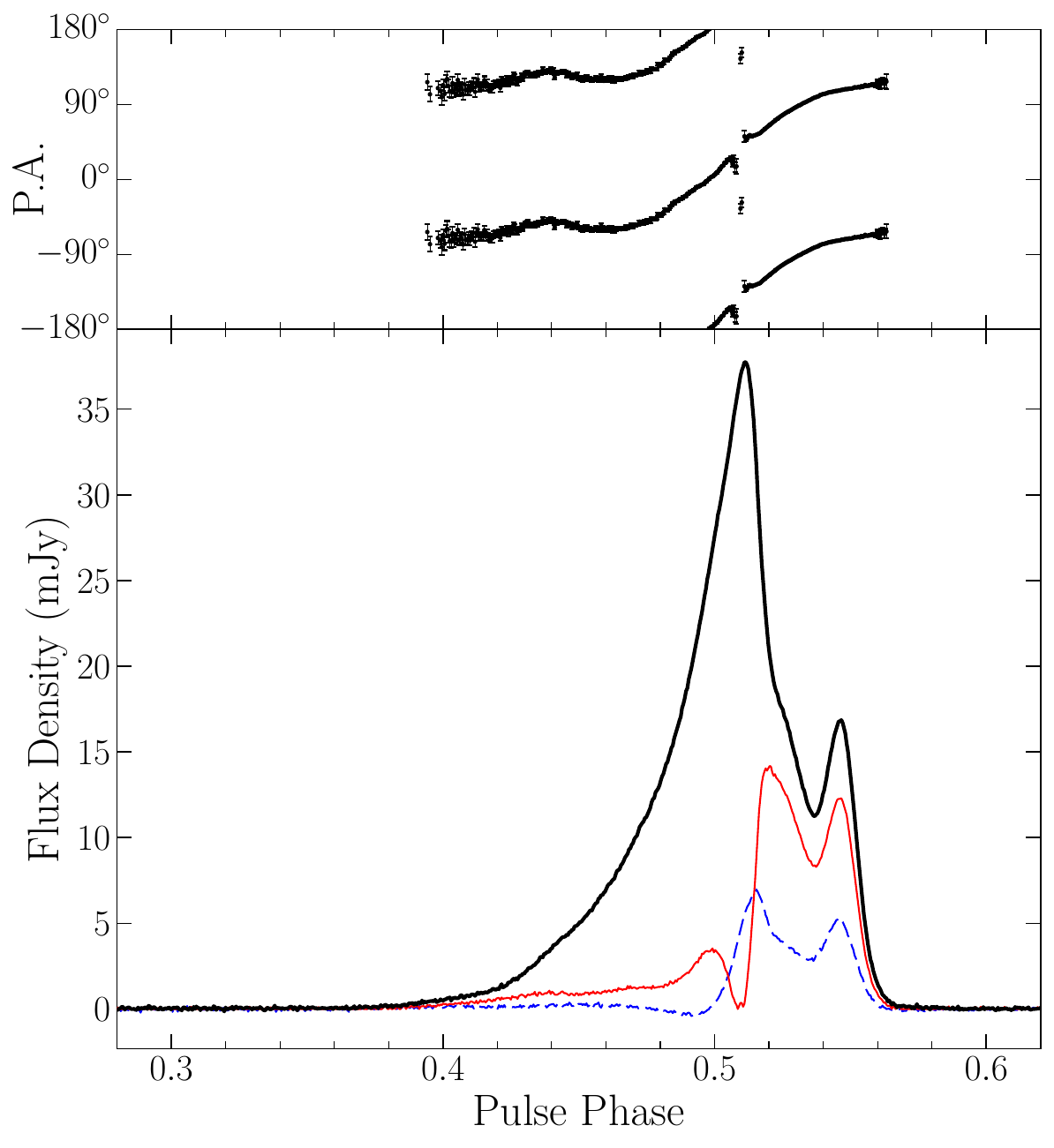}
    \caption{METM-calibrated polarization profile of PSR J1022+1001 at 2\,GHz. See Figure~\ref{fig:polprof430}'s caption for details. A Rotating Vector Model fit to the P.A.\ curve is also shown, and is described in the text.}
    \label{fig:polprofS}
\end{figure}

Each METM-calibrated, RM-corrected profile was added in phase using \texttt{psradd} to produce average polarization profiles for each receiver. These are shown in Figures~\ref{fig:polprof430}, \ref{fig:polprofL}, and \ref{fig:polprofS}, and the latter two include fitted RVM P.A.\ curves, described below.

In order to produce the 430-MHz profile shown in Figure~\ref{fig:polprof430}, we first corrected a sign flip in the P.A.\ and Stokes-$V$ curves, which were recorded as having the opposite sign of their L- and S-band counterparts. Considering the profiles of PSR J1022+1001 presented in \citet{stc+99}, which were obtained at Jodrell Bank, we see that the corresponding curves are broadly consistent in shape from 410 to 1414 MHz (a downward-sloping P.A.\ curve near the pulse peaks and a negative peak for $V$). That work uses the opposite sign convention for P.A.\ and $V$ from ours (P.A. increases clockwise and a left-handed signal has $V > 0$), and yet our 430-MHz P.A.\ and $V$ curves match their 410-MHz data. It is apparent that Arecibo's 430-MHz receiver used this convention during our observations, and this was not properly recorded in the \textsc{psrfits} file headers. We corrected for this error by flipping the signs of $U$ and $V$.

RMs, mean flux densities, and pulse widths are given in Table~\ref{tab:profiles-rvm}. We used \texttt{ionFR}\footnote{\url{https://github.com/csobey/ionFR}}, a package that estimates the ionospheric contribution to Faraday rotation \citep{ssh+13}, to ensure the reported RMs are purely astrophysical.

\begin{deluxetable*}{ccccccccc}
  \tabletypesize{\footnotesize}
  \tablewidth{\textwidth}
  \tablecaption{Rotation Measures, Flux Densities, Pulse Widths, and RVM Fits}
  \tablecolumns{9}
  \tablehead{
  \colhead{Frequency} & 
  \colhead{RM} &
  \colhead{Flux Density} &
  \colhead{$W_{10}$} &
  \colhead{$W_{50}$} &
  \colhead{\phtnm$\varphi_0$\tablenotemark{a}} &
  \colhead{$\alpha$} &
  \colhead{$\beta$} &
  \colhead{Reduced $\chi^2$} \\
  \colhead{(GHz)} &
  \colhead{(\rmu)} &
  \colhead{(mJy)} &
  \colhead{(ms)} &
  \colhead{(ms)} &
  \colhead{($\arcdeg$)} &
  \colhead{($\arcdeg$)} &
  \colhead{($\arcdeg$)} & 
  }
  \startdata
  0.43 & 8(2)\phn & 25(16) & 2.24 & 1.15 & \ldots & \ldots & \ldots & \ldots \\
  1.4 & \phnm12(5)\phn & 2(3) & 1.98  & 1.01 & $-$1.09(2)\phn & 9(1) & $-$1.1(1) & 1461 \\
  2.0 & \phnm\phn5(5)\phn & 2(3) & 1.82 & 0.42 & $-$0.405(8) & 6(6) & $-$0.6(6) & 201
  \enddata
  \tablenotetext{a}{Relative to phase of the leading peak of the profile.}
  \tablecomments{We report RMs, mean flux densities, and pulse widths at 10\% and 50\% peak intensity for calibrated profiles of PSR J1022+1001 at each frequency. The RMs shown are the interstellar RMs (calculated after accounting for ionospheric Faraday rotation). The uncertainties shown for RM and flux density are the epoch-to-epoch standard deviations. The other columns report best-fit RVM parameters given by \psrchive's \texttt{psrmodel} routine (along with the reported 1-$\sigma$ uncertainties). See Equation~\ref{eq:RVM} for a full description of the RVM parameters.}
  \label{tab:profiles-rvm}
\end{deluxetable*}

We applied the RVM (see Equation~\ref{eq:RVM}) to PSR J1022+1001's P.A.\ curve, using the \texttt{psrmodel} routine in \psrchive. Bins were only used for the fit if $L > 3\sigma_I$, where $\sigma_I$ is the standard deviation in $I$. The resulting best-fit parameters are also presented in Table~\ref{tab:profiles-rvm}. We were unable to reach a convergent fit for the 430\,MHz profile. 

Neither the 1.4 or 2\,GHz RVM fits are of particularly good quality, with very high reduced $\chi^2$. This is likely due, at least in part, to the ``notch'' seen in the P.A.\ curve near the phase when the beam is closest to the line of sight. The slope at this phase is used to estimate $\alpha$ and $\beta$:
\begin{equation} \label{eq:RVMII}
    \left.\frac{d\Psi}{d\varphi}\right\vert_{\varphi_0} = \frac{\sin\alpha}{\sin\beta}\,.
\end{equation}

\citet{stc+99} were able to obtain a better RVM fit with their 1.4\,GHz data from Jodrell Bank, finding $\alpha = 83(27)\arcdeg$ and $\beta = -7\fdg1(4)$. Notably, their P.A.\ curve lacks the ``notch'' present in ours, but this could be simply due to a lack of points in the given phase range.

The poor quality of our RVM fits and their disagreement with prior studies suggests that despite the P.A.\ curve's visual similarity to the well-behaved curves common for slower pulsars, PSR J1022+1001's emission geometry is more complicated than what may be easily described by the RVM. One such complex geometry was explored by \citet{rk+03}.

\subsection{Pulse Profile Variability} \label{sec:variability}

\begin{figure}
    \centering
    \includegraphics[width=1\linewidth]{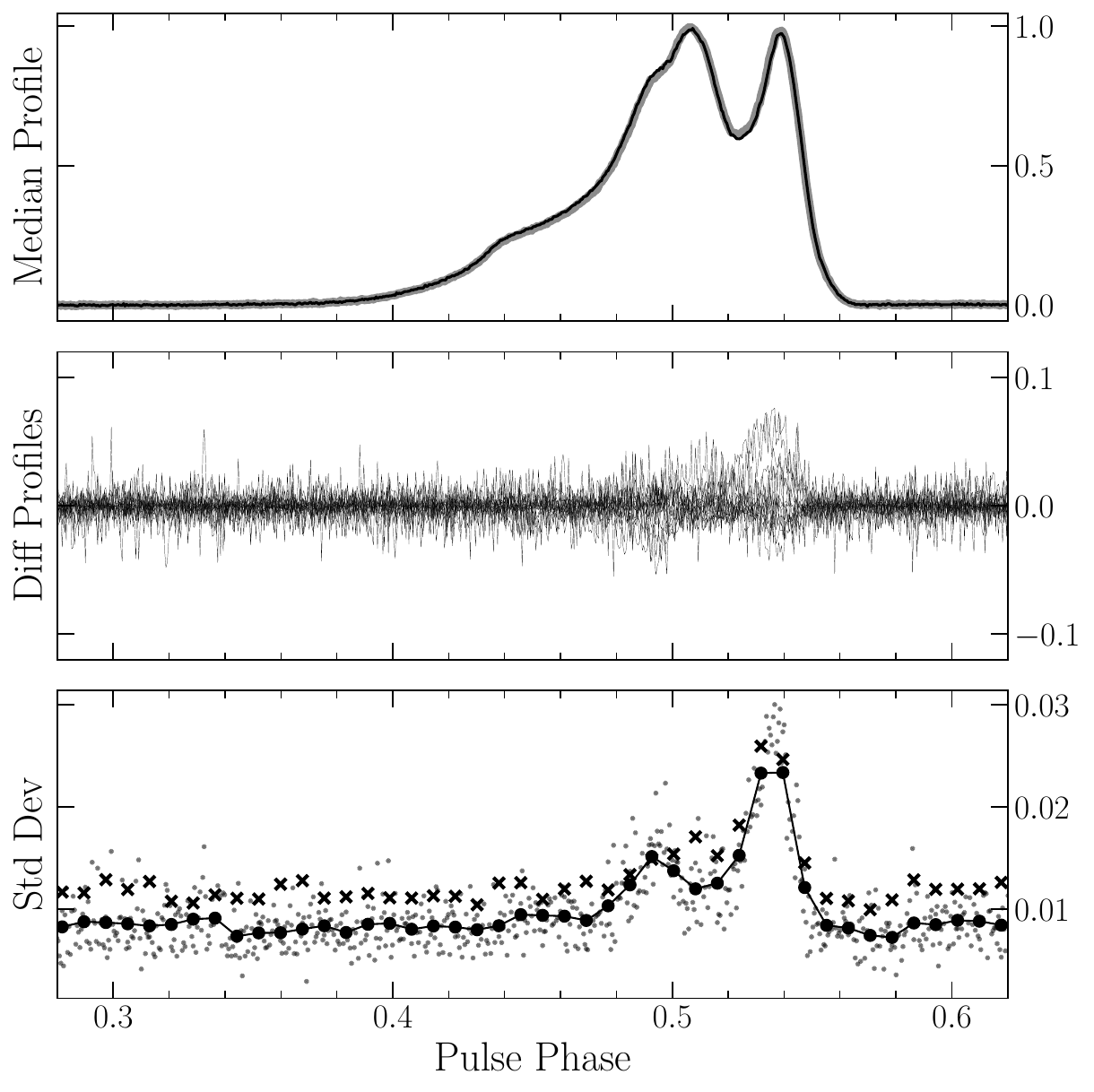}
    \caption{Profile variability visualization for 430\,MHz data. The top panel shows median pulse profiles from IFA- (thick gray line) and METM-calibrated (thin black line) data, the middle panel shows METM-calibrated difference profiles (data subtracted by the median profile), and the bottom panel shows the binned standard deviation of the difference profiles (small grey points), as well as a moving average with only 128 bins (large black points). The moving average is also shown with `X'-shaped markers for IFA-calibrated data.}
    \label{fig:variability430}
\end{figure}

\begin{figure}
    \centering
    % \vspace{5mm}
    \includegraphics[width=1\linewidth]{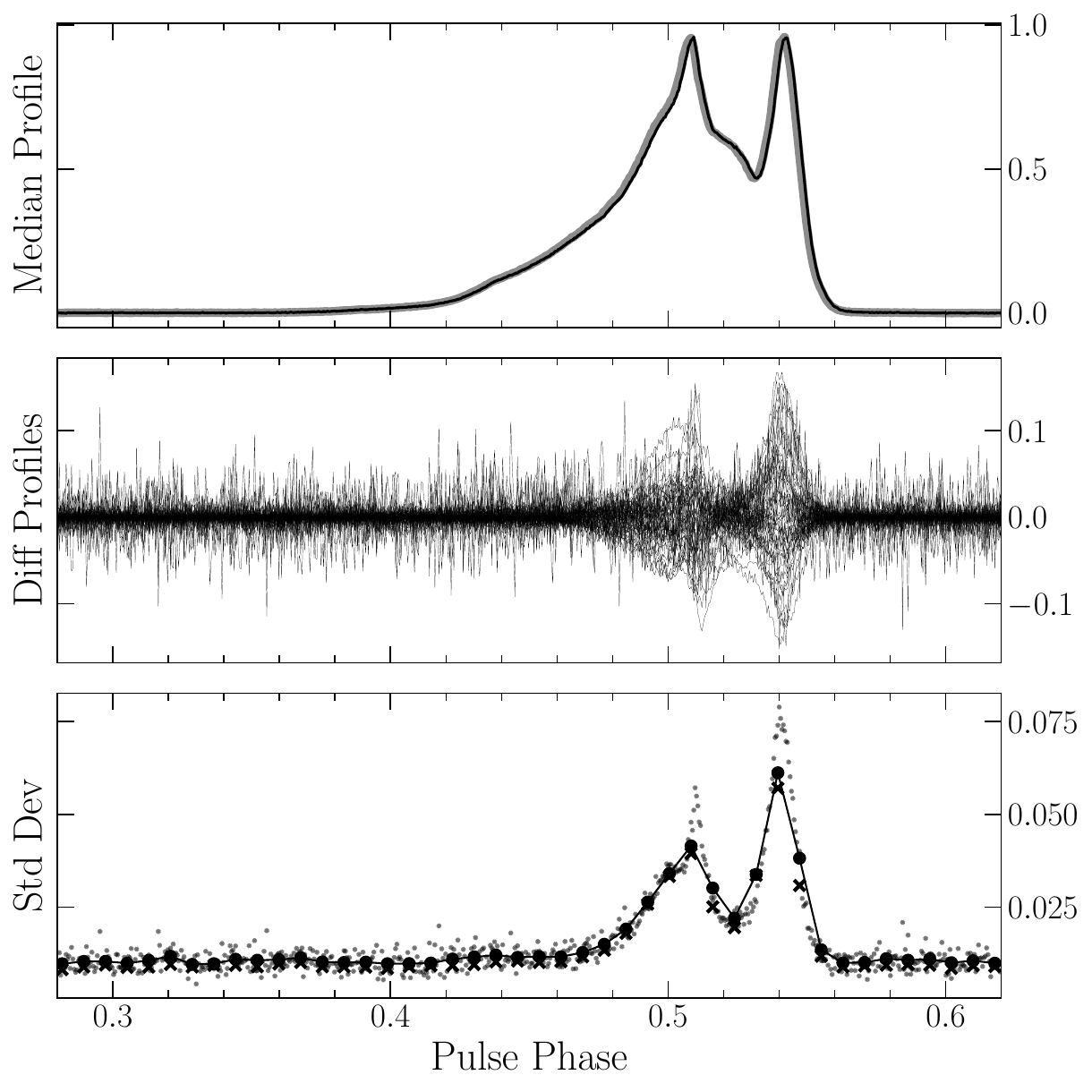}
    \caption{Profile variability visualization for 1.4\,GHz data. See Figure~\ref{fig:variability430}'s caption for details.}
    \label{fig:variabilityL}
\end{figure}

\begin{figure}
    \centering
    % \vspace{5mm}
    \includegraphics[width=1\linewidth]{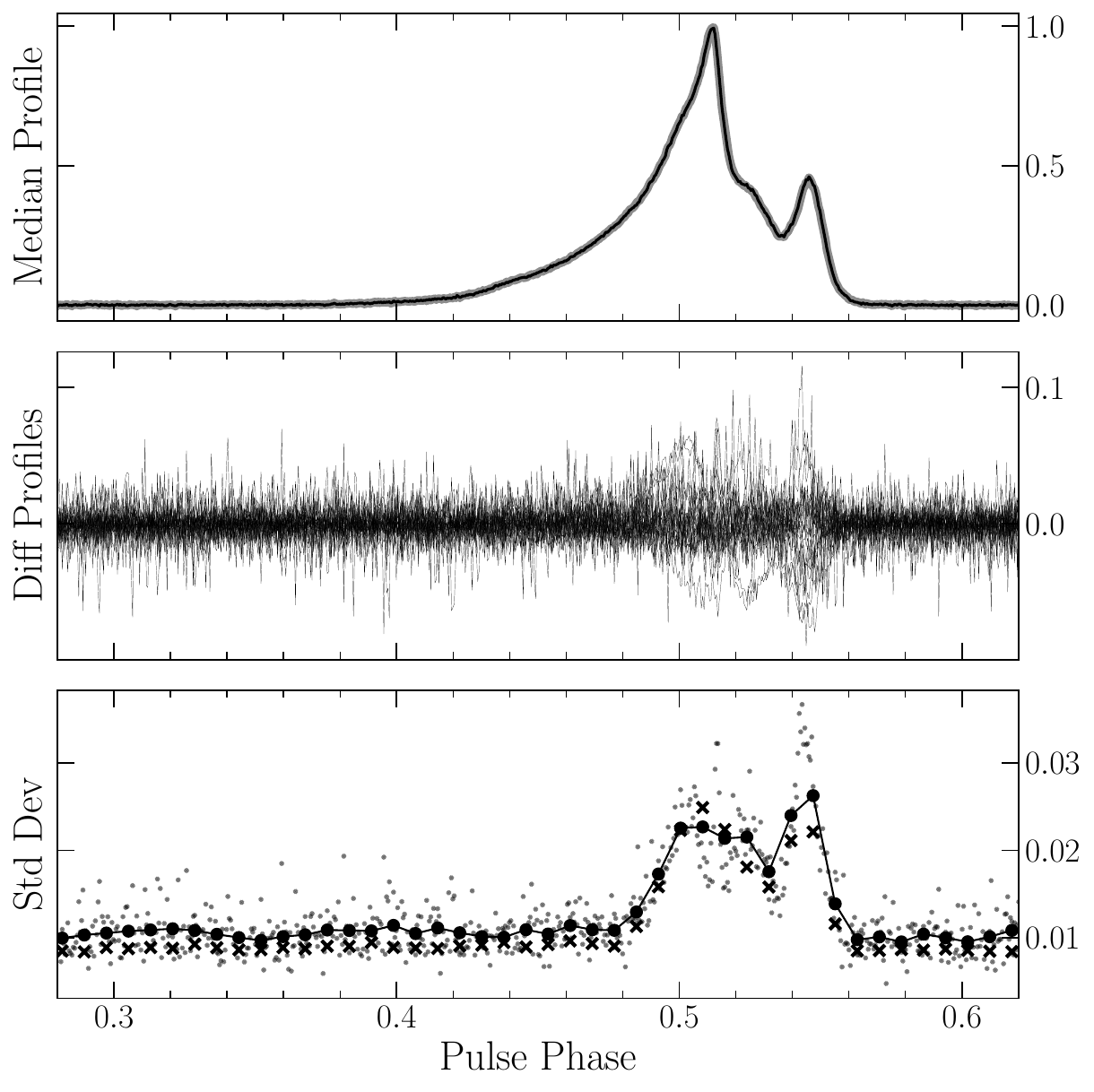}
    \caption{Profile variability visualization for 2\,GHz data. See Figure~\ref{fig:variability430}'s caption for details.}
    \label{fig:variabilityS}
\end{figure}

Figures \ref{fig:variability430}, \ref{fig:variabilityL}, and \ref{fig:variabilityS} show the profile variability of both the IFA and METM data sets in the style of \citet{hbo+04}, with the modifications introduced in \citet{pbc+21}. We phase-aligned the profiles with \texttt{pas}, then normalized each profile by scaling it such that its integral is equal to unity \citep[``flux-based normalization'' in][]{hbo+04}. In the normalization stage, profiles with S/N below a threshold value were removed: such profiles are problematic because their integral approaches zero. A S/N threshold was chosen separately for each receiver in order to balance keeping as many profiles as possible with removing the noisiest few. For each receiver, we created a comparison profile by calculating the median value of the normalized profiles in each phase bin. Then, we created difference profiles by subtracting the median profile from the individual profiles on each epoch. We then quantified the variability by calculating the standard deviation of the difference profile values in each phase bin. In each data-set/receiver combination, these binned standard deviations are noticeably higher near the peaks of the pulse profile.

Another way to characterize the variability of the pulse profile is to compute the relative phases and heights of the two peaks that characterize PSR J1022+1001's pulse profile. Following the `peak optimizer' method presented by \citet{pbc+21}, we estimated each peak's phase and height by fitting a parabola to the data within a certain number of bins of an initial guess of the peak phase bin. This guess was refined through multiple iterations.

\begin{figure}
    \centering
    \includegraphics[width=0.4\textwidth]{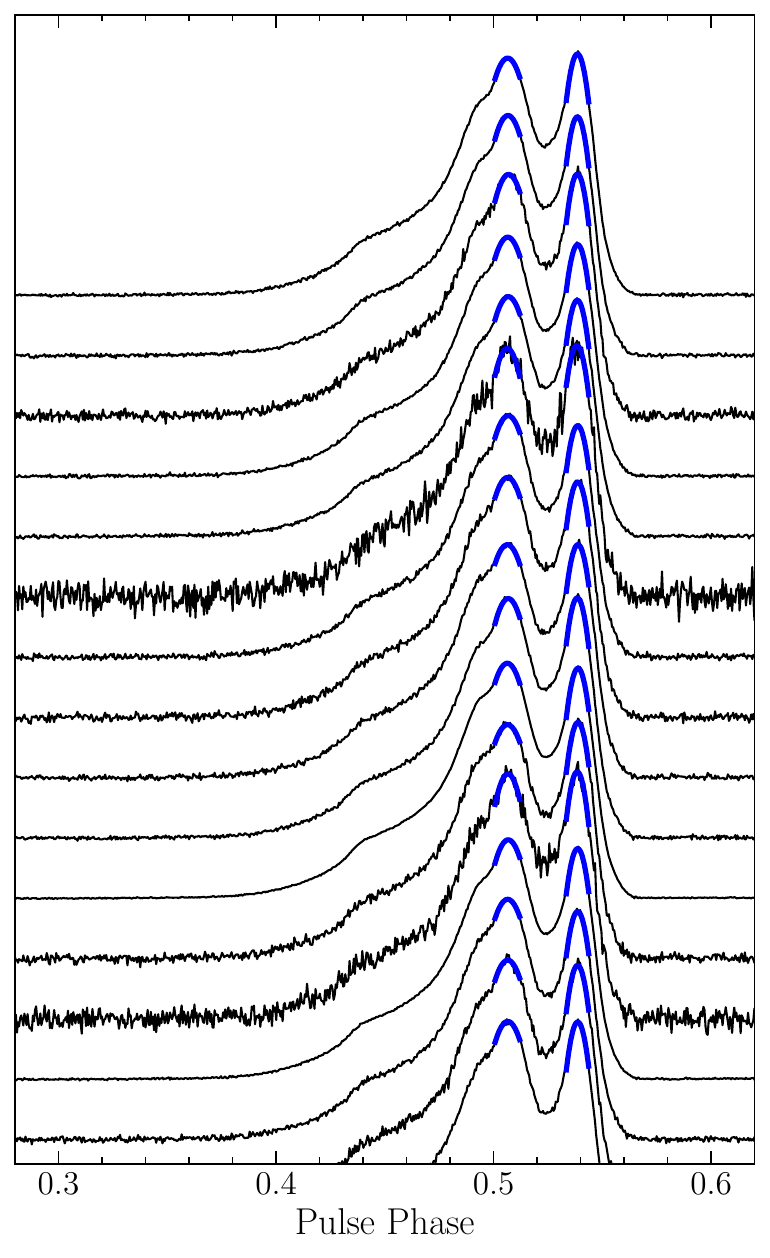}
    \caption{Best-fit parabolas (thick blue lines) used to estimate the phases and heights of the two peaks of each 430-MHz profile. The profiles are normalized such that their integrals equal the same (arbitrary) value, and they are stacked vertically simply to show each profile shape and polynomial fit.}
    \label{fig:waterfall_fits}
\end{figure}

\begin{figure}
    \centering
    \includegraphics[width=1\linewidth]{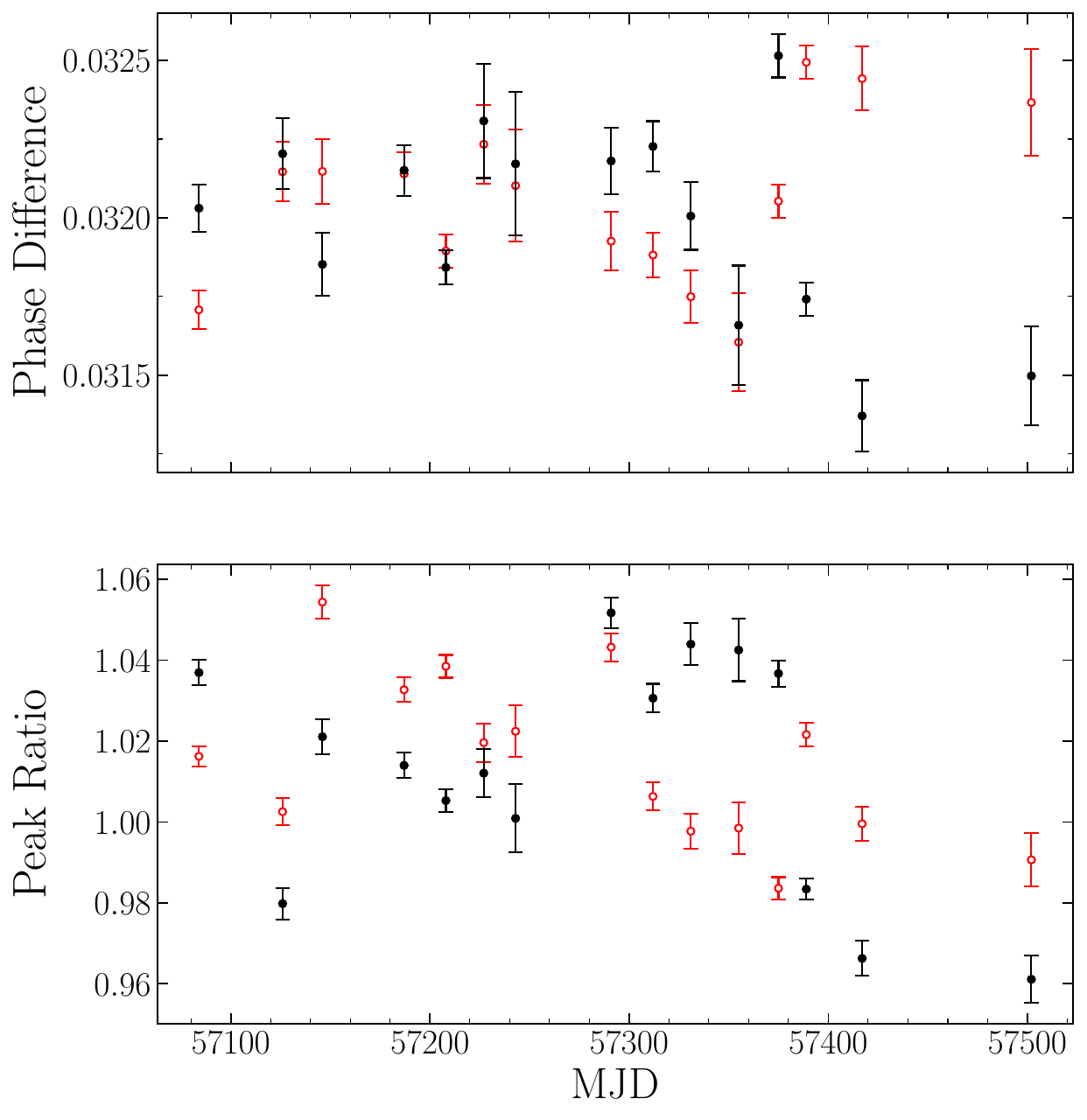}
    \caption{The results of the peak fitting method for 430\,MHz data. The top panel shows differences in pulse phase between the first and second peaks. The bottom panel shows peak ratios (first peak/second peak). Each panel shows results for METM-calibrated data with filled black points, and for IFA-calibrated data with open red points. Error bars show the 1-$\sigma$ uncertainties reported by the least-squares fit.}
    \label{fig:phases_ratios_430}
\end{figure}

\begin{figure}
    \centering
    \includegraphics[width=1\linewidth]{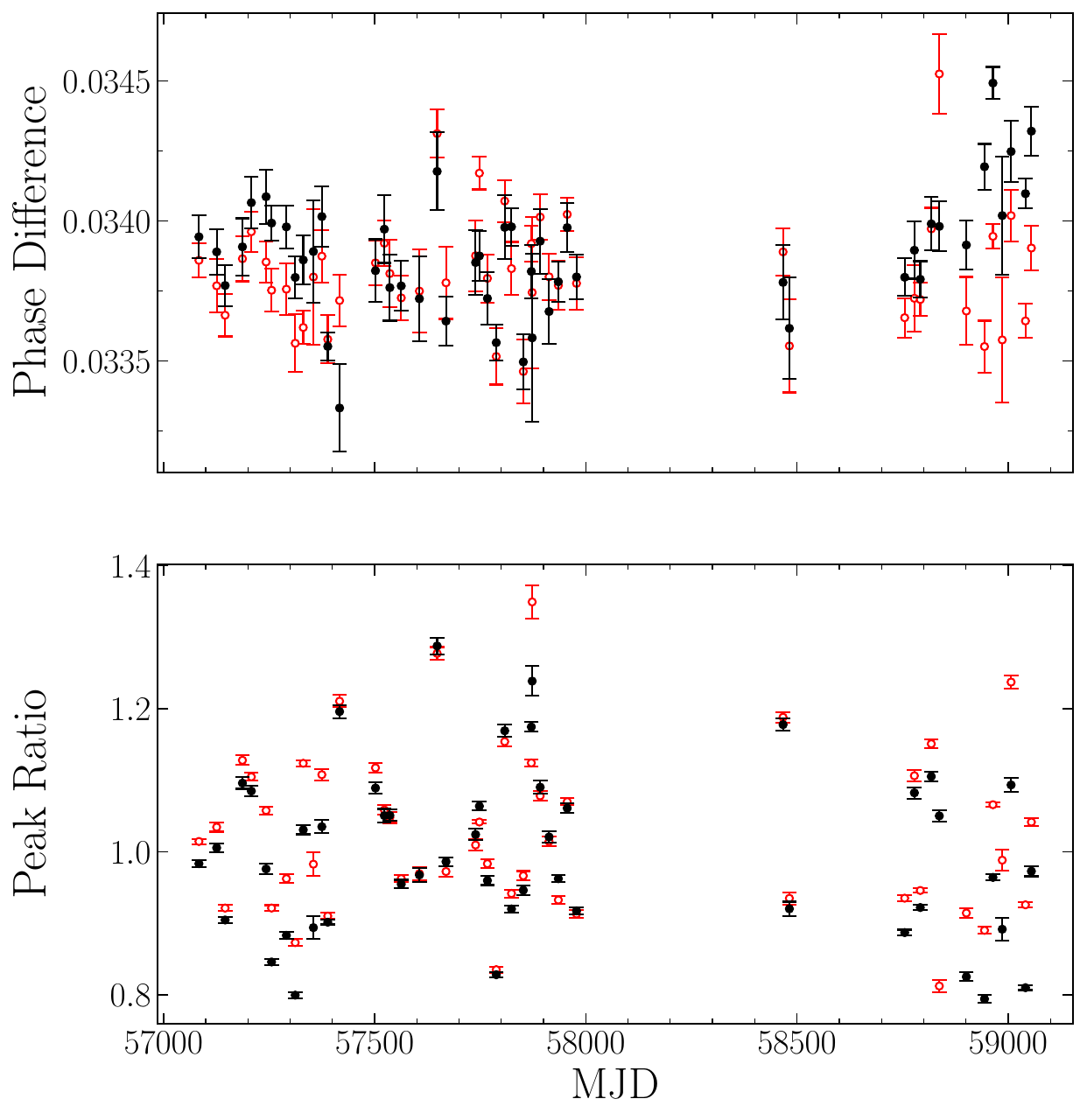}
    \caption{The results of the peak fitting method for 1.4\,GHz data. See Figure~\ref{fig:phases_ratios_430}'s caption for details.}
    \label{fig:phases_ratios_L}
\end{figure}

\begin{figure}
    \centering
    \includegraphics[width=1\linewidth]{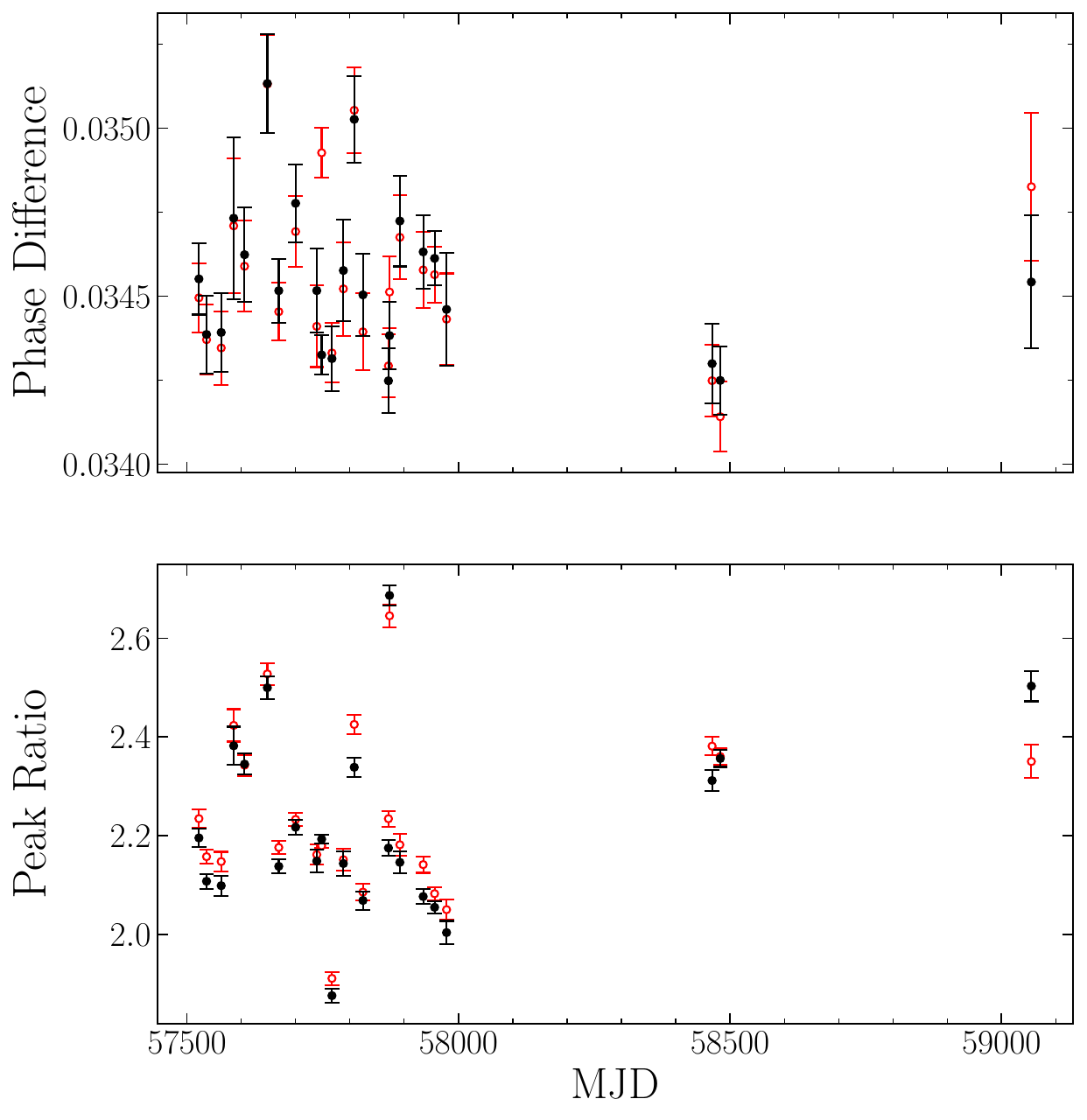}
    \caption{The results of the peak fitting method for 2\,GHz data. See Figure~\ref{fig:phases_ratios_430}'s caption for details.}
    \label{fig:phases_ratios_S}
\end{figure}

The results of the peak optimizer fit are shown in Figure~\ref{fig:waterfall_fits}. We take the turning point of the best-fit parabola to be the phase of the peak, and the height of the polynomial to be the peak height. The best-fit phase differences and peak height ratios are plotted in Figures \ref{fig:phases_ratios_430}, \ref{fig:phases_ratios_L}, and \ref{fig:phases_ratios_S}. 

\begin{figure*}
    \centering
    % \vspace{5mm}
    \includegraphics[width=1\textwidth]{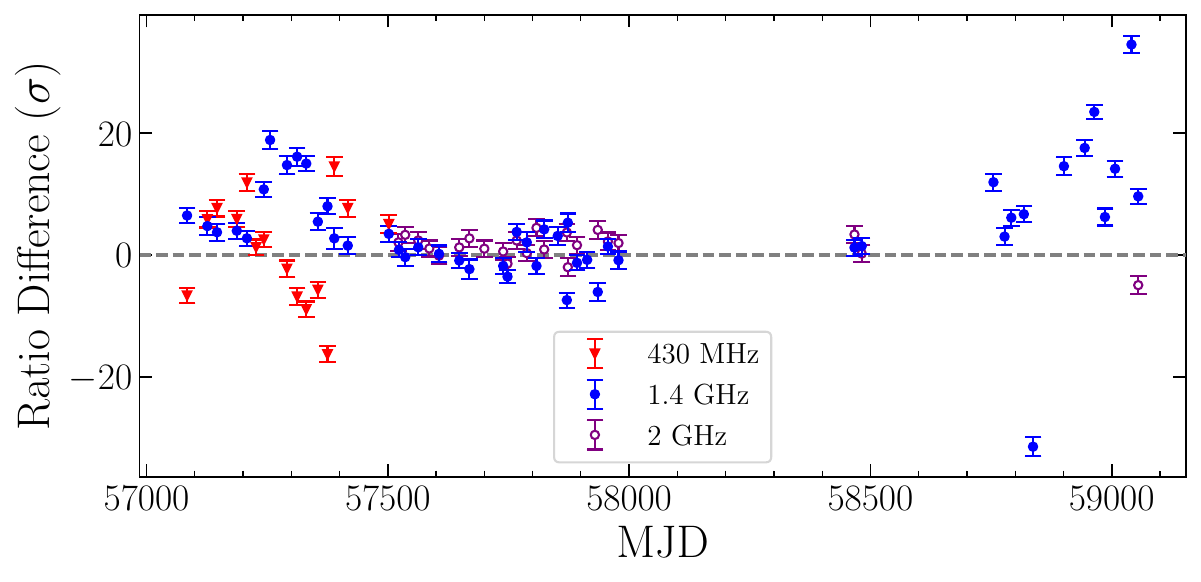}
    \caption{Each point, representing a single observation, shows the difference between the peak ratio obtained from IFA- and METM-calibrated data (i.e. corresponding red and black points from the bottom panels of figures \ref{fig:phases_ratios_430}, \ref{fig:phases_ratios_L}, and \ref{fig:phases_ratios_S}), in terms of the 1-$\sigma$ uncertainty on the METM-calibrated peak ratio. Red triangles are 430-MHz data, filled blue circles are 1.4\,GHz, and empty purple circles are 2\,GHz. The dashed gray line denotes equal peak ratios.}
    \label{fig:crossrcvr}
\end{figure*}

We investigated whether the choice of calibration technique influenced the average pulse shape by performing a Kolmogorov-Smirnov test to compare the peak ratios of the IFA and METM data sets from each receiver. In each case, the test statistic does not indicate that the two data sets are drawn from different distributions. The lowest $p$-value was 0.54, for the 1.4\,GHz data. As can be seen from Figure~\ref{fig:crossrcvr}, there are certainly examples of IFA- and METM-calibrated data from the same observation having different pulse shapes, there is no consistent long-term difference.

\begin{figure}
    \centering
    \includegraphics[width=1\linewidth]{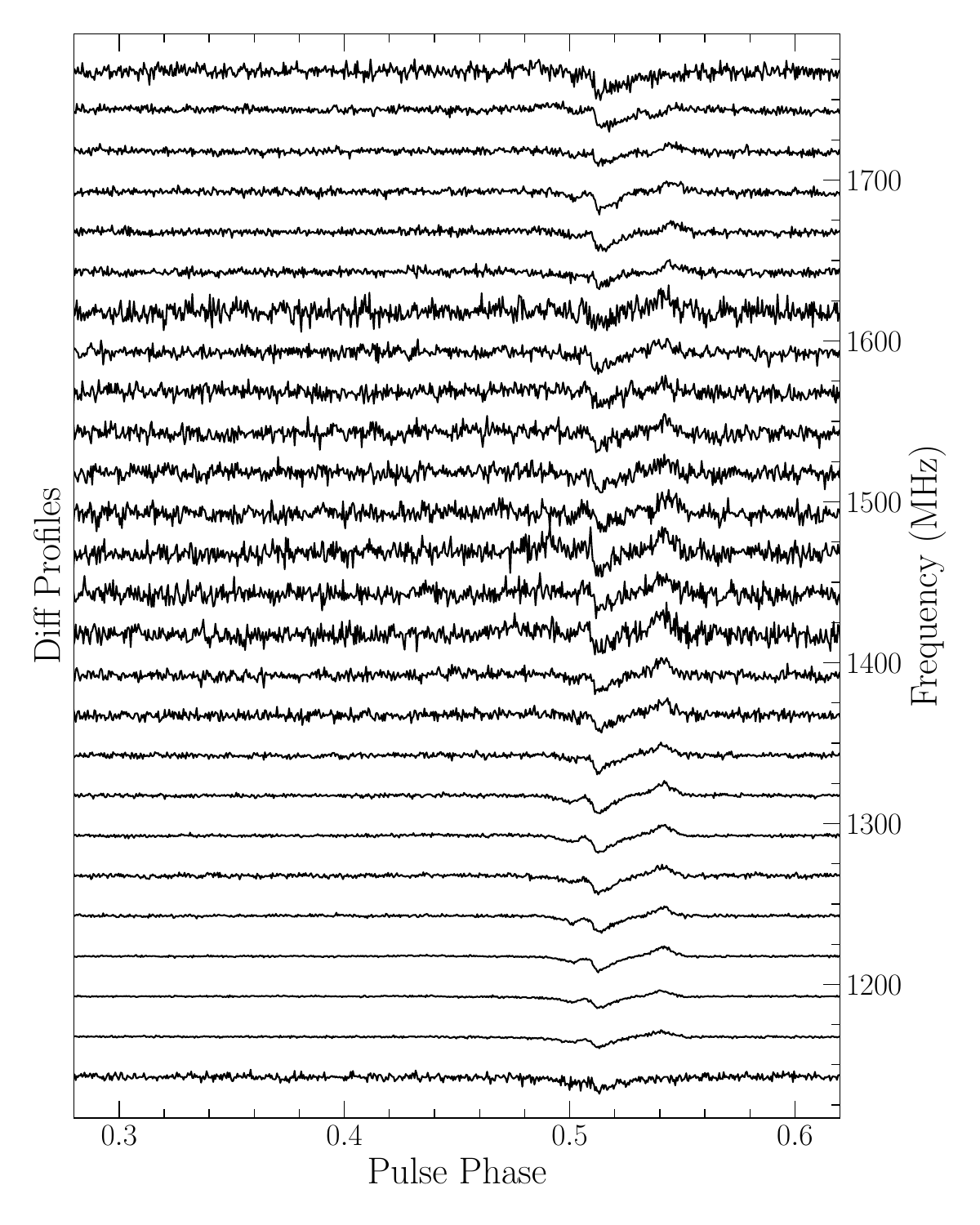}
    \caption{Example deviations from the median pulse shape in each 25-MHz subband, from a 1.4-GHz observation on 24 April 2016 (MJD 57502).}
    \label{fig:subbands}
\end{figure}

When coupled with profile frequency evolution, interstellar scintillation could potentially cause changes in the profile shape. 
In our 1.4-GHz data, scintillation bandwidths range from 40--100\,MHz and timescales are $\sim$20\,min (these are merely rough estimates based on visual inspection of our dynamic spectra).
%We estimate J1022's scintillation bandwidth to be 98\,MHz at 1.4\,GHz, using \texttt{NE2001p}\footnote{\url{https://github.com/stella-ocker/mwprop}} \citep{oc+24} and Equation 10 of \citet{cl+02}. 
To investigate this possibility, we created profiles from 25-MHz bandwidths. For each subband, we created a comparison median profile in the same way as in the \citet{hbo+04} method described above, so that difference profiles show the deviation from the actual pulse shape at a particular frequency, without being biased by profile frequency evolution. We found (see Figure~\ref{fig:subbands} for an example) these subbanded profiles exhibited the same variability across the band within each observation, which cannot be explained by scintillation.

\begin{figure}
    \centering
    \includegraphics[width=1\linewidth]{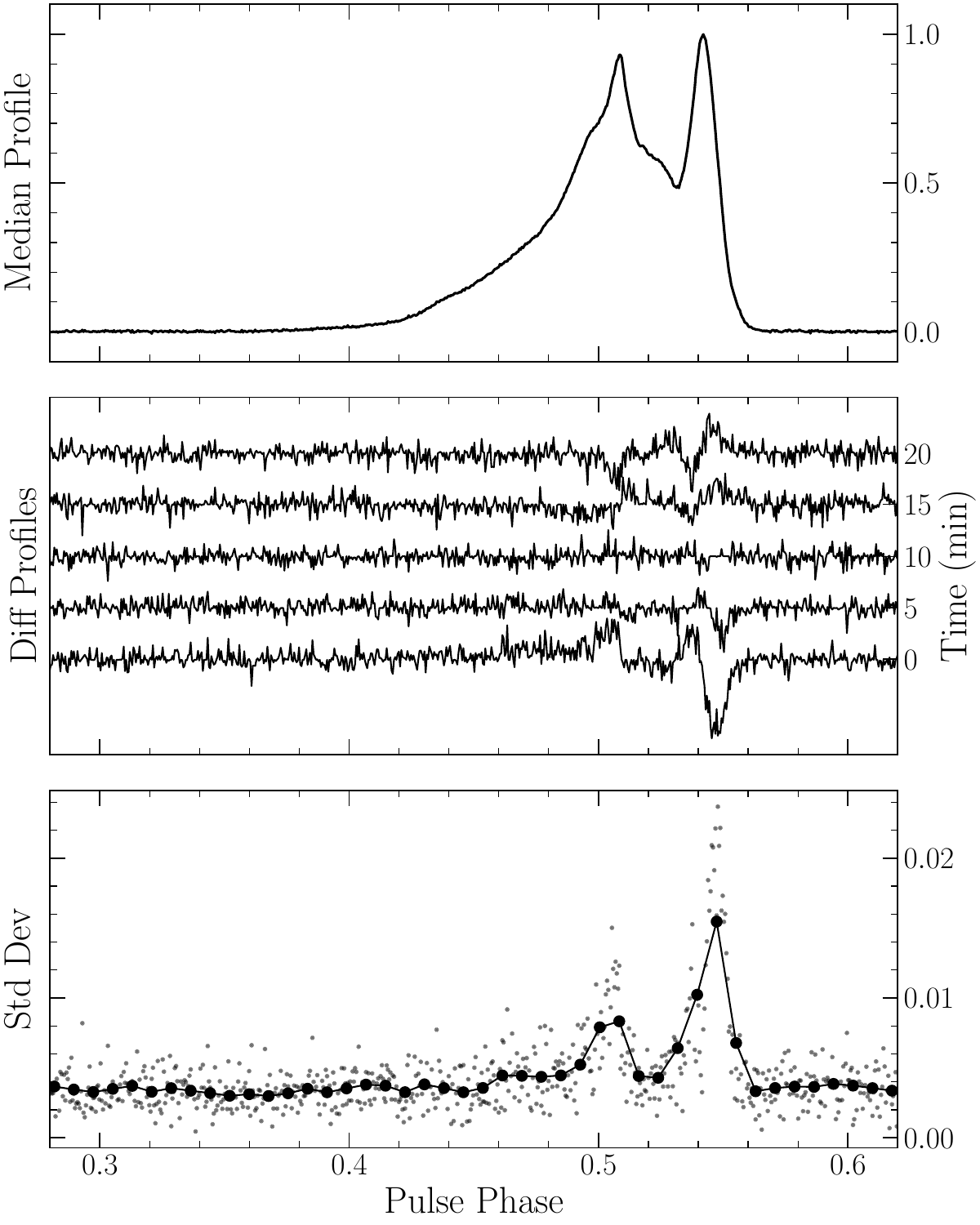}
    \caption{Example of an observation that shows changing variability between 5-minute subintegrations. The middle panel shows a difference profile from each subintegration: the data minus the median profile, which is shown in the top panel. The bottom panel shows the binned standard deviations in the same manner as in Figure \ref{fig:variability430}. Data are from a 1.4-GHz observation made on 3 March 2015 (MJD 57084).}
    \label{fig:subints}
\end{figure}

We also looked for variability within individual observations. For each observation at least 20 minutes long, we split the data into 5-minute subintegrations and subjected the resulting profiles to the \citet{hbo+04} method. 
In most observations, comparing the subintegration profiles did not yield any signs of variability over the course of the observation itself.
However, in a few cases the pulse shape did change, seemingly smoothly over a $\sim$10--15 minute timescale: see Figure~\ref{fig:subints} for an example of this.
The possibility that this short-timescale variability in particular is due to scintillation coupled with the frequency-dependent profile shape, as mentioned above, has not been ruled out.

\begin{deluxetable*}{l c c}
\centering
\tabletypesize{\footnotesize}
\tablewidth{0pt}
\tablecolumns{3}
\tablecaption{Timing Solution and Derived Parameters for PSR J1022+1001}
\label{tab:timing_solution}
\tablehead{
    \multicolumn{1}{l}{Parameter} &
    % \colhead{Parameter} &
    \colhead{METM data set} &
    \colhead{IFA data set}
    }
\startdata
\cutinhead{Timing Parameters}
Ecliptic longitude ($\arcdeg$) & 153.86583002(6) & 153.86583001(5) \\
Ecliptic latitude ($\arcdeg$) & $-$0.06380(6) & $-$0.06384(5) \\
Spin frequency (Hz) & 60.7794488569342(4) & 60.7794488570170(4) \\
Frequency derivative (Hz\,s$^{-1}$) & $-$1.6008(2)$\times 10^{-16}$ & $-$1.6011(2)$\times 10^{-16}$ \\
Dispersion measure (\dmu) & 10.263 & 10.266 \\
Proper motion in lon.\ (mas\,yr$^{-1}$) & $-$15.9(1) & $-$16.1(1) \\
Proper motion in lat.\ (mas\,yr$^{-1}$) & $-$2(1)$\times 10^{2}$ & $-$2(1)$\times 10^{2}$ \\
Parallax (mas) & 2.2(9) & 3.0(7) \\
Epoch (MJD) & 58048 & 58042 \\
Span (MJD) & 57041--59055 & 57028--59055 \\
Number of TOAs & 3822 & 3930 \\
Epoch-averaged WRMS (\us) & 3.49 & 3.36 \\
\cutinhead{Orbital Parameters}
Orbital period (days) & 7.8051301610(5) & 7.8051301612(4) \\
Projected semimajor axis (lt-s) & 16.7654214(5) & 16.7654227(5) \\
Eccentricity & 9.726(7)$\times 10^{-5}$ & 9.727(6)$\times 10^{-5}$ \\
Epoch of periastron (MJD) & 58051.8516(8) & 58044.0474(7) \\
Longitude of periastron ($\arcdeg$) & 97.91(4) & 97.95(3) \\
Projected semimajor axis derivative (lt-s\,s$^{-1}$) & \ldots & 3.5(9)$\times 10^{-14}$ \\
\cutinhead{Profile Frequency-Dependence Parameters}
FD1 (\us) & \phnm1.0(2)$\times 10^{-4}$ & \phnm1.3(1)$\times 10^{-4}$ \\
FD2 (\us) & $-$1.6(2)$\times 10^{-4}$ & $-$2.0(1)$\times 10^{-4}$ \\
FD3 (\us) & \phnm7.4(8)$\times 10^{-5}$ & \phnm8.7(7)$\times 10^{-5}$ \\
\cutinhead{Derived Parameters}
Spin period (ms) & 16.4529297123745(1) & 16.4529297123521(1) \\
Period derivative & 4.3333(5)$\times 10^{-20}$ & 4.3342(4)$\times 10^{-20}$ \\
Right ascension (J2000) & $10^\textrm{h}\, 22^\textrm{m}\, 57\, \fs218(5)$ & $10^\textrm{h}\, 22^\textrm{m}\, 57\, \fs214(4)$ \\
Declination (J2000) & $+10\arcdeg\, 01\arcmin\, 57\, \farcs7(2)$ & $+10\arcdeg\, 01\arcmin\, 57\, \farcs6(2)$ \\
Galactic longitude ($\arcdeg$) & 231.79000(5) & 231.79004(5) \\
Galactic latitude ($\arcdeg$) & 51.09864(4) & 51.09861(4) \\
Binary mass function (\Msun) & 8.31$\times 10^{-2}$ & 8.31$\times 10^{-2}$ \\
Minimum companion mass (\Msun) & 0.72 & 0.72
\enddata
\tablecomments{Numbers in parentheses are the 1-$\sigma$ uncertainties in the last digit as reported by \pint. We used the DD binary timing model, DE440 Solar System Ephemeris, the TT(BIPM2021) time standard, and the IERS2010 convention for ecliptic coordinates. 
The minimum companion mass assumes 
%a neutron star with  radius $R = 10$\,km, 
a pulsar mass of $1.4$\,M$_\odot$
%, and moment of inertia $I = 10^{45}$\;g\,cm$^2$. 
and an edge-on orbit, $i = 90\arcdeg$.
Not shown are the two fitted \texttt{JUMP} parameters, which account for the phase offsets between the three receivers at AO; and the 54/55 DMX parameters, for which the reported DMs are reference values.
}
\end{deluxetable*}

\subsection{Timing Solution Comparison} \label{sec:timingsolution}

The timing solution we obtained from the METM data set is shown in Table~\ref{tab:timing_solution} along with the timing solution calculated from the IFA-calibrated profiles. One noticeable difference to the IFA timing solution (which has only minor differences from the NG15 solution, caused by the removal of FD parameters, as discussed in \S\,\ref{sec:timing}) is the lack of a significant time derivative of the projected semimajor axis. We attribute this to the reduced S/N and fewer TOAs in the METM data set compared to the IFA data set, in which the parameter has only marginal (3.9-$\sigma$) significance. The reason for this is discussed below.

Table~\ref{tab:timing_comparison} shows various metrics of evaluating the suitability of both data sets' timing solutions. For 430\,MHz data, the METM method causes a significant improvement in each RMS value. Notably, there are only 158 TOAs at 430\,MHz in the METM data set, vs.\ 219 in the IFA data set. Differences in the number of TOAs, and S/N, result from frequency channels being eliminated due to RFI in the observations of PSRs B0525+21, B1937+21, or J0030+0451 (or the corresponding noise diode observations), or \texttt{pcm} failing to achieve a solution for the PR in certain frequency channels. In these ways, polarization calibration can corrupt frequency channels, leading to fewer (in the cases were subbands are eliminated altogether) or lower-S/N (when a limited number of frequency channels are corrupted within a subband) TOAs.

For 1.4 and 2\,GHz data, which have many more TOAs than the 430\,MHz data, IFA leads to a better timing solution than METM by every metric. Therefore, it would seem that calibrating NANOGrav observations using METM generally does not lead to an improvement in the timing solution over the standard IFA method.

\section{Summary \& Discussion} \label{sec:discussion}

PSR J1022+1001's pulse profile exhibits significant variability in the NANOGrav 15-yr data set, in data taken from 430\,MHz to 2\,GHz. We have applied a robust polarization calibration scheme to these data and see no increased stability as a result. These results agree with those of \citet{kxc+99} and \citet{pbc+21}, and run counter to \citet{hbo+04}, in finding significant pulse profile variability that does not appear to be caused by incorrect polarization calibration.

Our polarization calibration scheme does not lead to an overall improvement in the quality of the timing solution obtained from the data. In fact, data that only underwent a standard polarization calibration procedure have a slightly higher-quality fit. These results are consistent with a parallel study that more thoroughly compares NANOGrav timing results between different polarization calibration methods using data from the Green Bank Telescope \citep{dmw+24}. We may therefore tentatively conclude that incorrect polarization calibration is not responsible for the pulse profile changes observed, though it is of course possible that yet more accurate polarization calibration could be explored.

Because the pulse profiles obtained using narrower bandwidths are variable in the same way as frequency-averaged profiles, a potential cause in the combination of profile frequency-dependence and interstellar scintillation, as proposed by \citet{sy+16}, is ruled out. This leaves mechanisms intrinsic to the pulsar as the most likely cause of the variability. As discussed in \S\ref{sec:intro}, these mechanisms include pulse-to-pulse jitter, precession, and mode-changing. The long timescale of the variations is inconsistent with jitter, and the seemingly random variations on long (months to years) timescales are inconsistent with expectations for precession.
Single pulse studies have not indicated that PSR J1022+1001 exhibits nulling, mode-changing, or giant pulses, and measurements of jitter noise have been inconsistent \citep{sy+16,ppm+21,fhl+21}. 

While our observed pulse shapes do not appear to fall into a number of discrete modes, it is possible that modeling the pulse shape as linear combinations of averages of the most extreme peak-ratio profiles could lead to evidence of mode-changing, as it did for PSR B1828$-$11 \citep{slk+19}.
Aside from mode-changing, which merits further investigation along the lines of PSR B1828$-$11, the explanation for the observed variability that remains is some phenomenon in the pulsar magnetosphere, such as those proposed by \citet{rk+03} and \citet{lkl+15}.

PSR J1022+1001 has respectable timing precision despite its variable pulse shape: its epoch-averaged rms residual in the NANOGrav 15-yr data set is about 3\,\us. One potential way to improve precision further is by accounting for the variability with a TOA generation method that allows pulse shapes to vary. While such methods have been demonstrated \citep{kxc+99,sc+12}, applying one in a PTA context would require careful consideration, and is beyond the scope of this analysis.

\begin{deluxetable*}{ccccccccc}
\centering
\tabletypesize{\footnotesize}
\tablewidth{0pt}
\tablecolumns{10}
\tablecaption{Timing Analysis Comparison \label{tab:timing_comparison}}
\tablehead{
    \colhead{Rcvr} &
    \colhead{Method} &
    \colhead{$N_\textrm{TOA}$} &
    \colhead{S/N$_\textrm{med}$} &
    \colhead{RMS$_\textrm{all}$} &
    \colhead{RMS$_\textrm{avg}$} &
    \colhead{WRMS$_\textrm{all}$} &
    \colhead{WRMS$_\textrm{avg}$} &
    \colhead{EFAC} \\
    &
    &
    &
    &
    (\us) &
    (\us) &
    (\us) &
    (\us) & 
    }
\startdata
\multirow{2}{*}{All} 
& IFA & 3930 & 148.5 & 4.53 & 6.37 & 2.68 & 3.36 & \ldots \\ 
& METM & 3822 & 143.8 & 3.79 & 7.25 & 2.81 & 3.49 & \ldots \\ 
\hline 
\multirow{2}{*}{430} 
& IFA & 219 & 251.4 & 9.90 & 11.47\phn & 9.91 & 10.07\phn & 1.26 \\ 
& METM & 158 & 249.8 & 4.94 & 8.03 & 4.96 & 6.51 & 1.30 \\ 
\hline 
\multirow{2}{*}{L-wide} 
& IFA & 2559 & 163.2 & 2.17 & 5.33 & 2.19 & 2.91 & 1.07   \\ 
& METM & 2555 & 161.8 & 2.24 & 5.67 & 2.26 & 3.01 & 1.14 \\ 
\hline 
\multirow{2}{*}{S-wide} 
& IFA & 1152 & 115.2 & 3.33 & 7.08 & 3.27 & 4.40 & 1.14 \\ 
& METM & 1109 & 101.0 & 4.91 & 9.90 & 4.85 & 5.19 & 1.29
\enddata
\tablecomments{Timing analysis comparison between the IFA and METM data sets. Columns report the number of TOAs ($N_\textrm{TOA}$), median TOA S/N (S/N$_\textrm{med}$), unweighted (RMS) and weighted (WRMS) root mean square of all or epoch-averaged TOAs. The first set of two rows shows results for all combined data, and subsequent rows show results for data from individual receivers, as well as the EFAC white noise parameter for each receiver.}
\end{deluxetable*}

\section*{Author Contributions}

W.F.\ performed the analysis and prepared the text, figures, and tables. M.A.M.\ provided valuable guidance and feedback over the course of the project. The polarization calibration procedure was based on previous work by P.A.G.\ and H.M.W., and was further refined by W.F.\ and L.D. K.C. and I.H.S.\ ran the observations of PSR B0525+21. G.A., A.A., A.M.A., Z.A., P.T.B., P.R.B., H.T.C., K.C., M.E.D., P.B.D., T.D., E.C.F, W.F., E.F., G.E.F., N.G.D., P.A.G., J.G., D.C.G., J.S.H., R.J.J., M.L.J., D.L.K., M.K., M.T.L., D.R.L., J.L., R.S.L., A.M., M.A.M., N.M., B.W.M., C.N., D.J.N., T.T.P., B.B.P.P., N.S.P., H.A.R., S.M.R., P.S.R., A.S., C.S., B.J.S., I.H.S., K.S., A.S., J.K.S., and H.M.W.\ developed the NANOGrav 15-yr data set through a combination of observations, data reduction, code development, timing analysis, and other contributions detailed in \citet{NG15data}.

\section*{Acknowledgements}
%\begin{acknowledgments}
The authors thank Loren Anderson, who, along with M.A.M., D.L.K., and D.R.L., gave valuable comments on this paper as part of reviewing W.F.'s Ph.D.\ thesis.

The NANOGrav Collaboration receives support from National Science Foundation (NSF) Physics Frontiers Center award Nos.\ 1430284 and 2020265, the Gordon and Betty Moore Foundation, NSF AccelNet award No.\ 2114721, an NSERC Discovery Grant, and CIFAR. The Arecibo Observatory is a facility of the NSF operated under cooperative agreement (AST-1744119) by the University of Central Florida (UCF) in alliance with Universidad Ana G.\ M\'{e}ndez (UAGM) and Yang Enterprises (YEI), Inc.

M.A.M.\ is supported by NSF \#2009425.
M.A.M.\ and D.R.L.\ are supported by NSF \#1458952.
P.R.B.\ is supported by the Science and Technology Facilities Council, grant number ST/W000946/1.
Pulsar research at UBC is supported by an NSERC Discovery Grant and by CIFAR.
K.C.\ is supported by a UBC Four Year Fellowship (6456).
M.E.D.\ acknowledges support from the Naval Research Laboratory by NASA under contract S-15633Y.
L.D.\ is supported by a WVU postdoctoral fellowship.
T.D.\ and M.T.L.\ are supported by an NSF Astronomy and Astrophysics Grant (AAG) award number 2009468.
E.C.F.\ is supported by NASA under award number 80GSFC21M0002.
G.E.F.\ is supported by NSF award PHY-2011772.
The Dunlap Institute is funded by an endowment established by the David Dunlap family and the University of Toronto.
T.T.P.\ acknowledges support from the Extragalactic Astrophysics Research Group at E\"{o}tv\"{o}s Lor\'{a}nd University, funded by the E\"{o}tv\"{o}s Lor\'{a}nd Research Network (ELKH), which was used during the development of this research.
N.S.P.\ was supported by the Vanderbilt Initiative in Data Intensive Astrophysics (VIDA) Fellowship.
H.A.R.\ is supported by NSF Partnerships for Research and Education in Physics (PREP) award No.\ 2216793.
S.M.R.\ and I.H.S.\ are CIFAR Fellows.
Portions of this work performed at NRL were supported by ONR 6.1 basic research funding.
%\end{acknowledgments}

\facilities{AO (PUPPI)}

\software{\texttt{astropy} \citep{astropy}, \textsc{enterprise} \citep{enterprise}, \texttt{ionFR} \citep{ssh+13}, \texttt{matplotlib} \citep{matplotlib},
% \texttt{NE2001p} \citep{cl+02,oc+24}, 
\texttt{numpy} \citep{numpy}, \pint \citep{pint+21,pint+24}, \psrchive \citep{psrchive}, \texttt{scipy} \citep{scipy}}

% \appendix

% \section{Appendix A}

% Appendix A

% \section{Appendix B}

% Appendix B

\bibliography{main}{}
\bibliographystyle{aasjournal}

\end{document}